\documentclass{aastex62}

\usepackage{amssymb,amsmath}

\shorttitle{Cataclysmic Variables}
\shortauthors{Hou et al.}

\begin{document}

\title{Spectroscopically Identified Cataclysmic Variables from LAMOST survey. I. The Sample}

\correspondingauthor{A-Li Luo}
\email{lal@nao.cas.cn, whou@nao.cas.cn}

\author{Wen Hou}
\affil{Key Laboratory of Optical Astronomy, National Astronomical Observatories, Chinese Academy of Sciences, \\
20A Datun Road, Chaoyang District, Beijing, 100101, China}

\author[0000-0001-7865-2648]{A-li Luo}
\affiliation{Key Laboratory of Optical Astronomy, National Astronomical Observatories, Chinese Academy of Sciences, \\
20A Datun Road, Chaoyang District, Beijing, 100101, China}
\affiliation{University of Chinese Academy of Sciences, Beijing 100049, China}
\affiliation{School of Information Management $\&$ Institute for Astronomical Science, Dezhou University, Dezhou 253023, China}
\affiliation{Department of Physics and Astronomy, University of Delaware, Newark, DE 19716, USA}

\author{Yin-Bi Li}
\affiliation{Key Laboratory of Optical Astronomy, National Astronomical Observatories, Chinese Academy of Sciences, \\
20A Datun Road, Chaoyang District,  Beijing, 100101, China}

\author{Li Qin}
\affiliation{Key Laboratory of Optical Astronomy, National Astronomical Observatories, Chinese Academy of Sciences, \\
20A Datun Road, Chaoyang District, Beijing, 100101, China}
\affiliation{University of Chinese Academy of Sciences, Beijing 100049, China}
\affiliation{School of Information Management $\&$ Institute for Astronomical Science, Dezhou University, Dezhou 253023, China}

\begin{abstract}

A sample of Cataclysmic Variables (CVs) is presented including spectroscopically identified 380 spectra of 245 objects, of which 58 CV candidates are new discoveries. The BaggingTopPush and the Random Forest algorithms are applied to the Fifth Data Release (DR5) of LAMOST to retrieve CVs with strong emission lines and with broad absorption lines respectively. Based on spectroscopic classification, 134 dwarf novae, 41 nova-like variables and 19 magnetic CVs are identified from the sample. In addition, 89 high--inclination systems and 33 CVs showing companion stars are recognized and discussed for their distinct spectral characteristics. Comparisons between CVs from LAMOST and from published catalogs are made in spatial and magnitude distribution, and the difference of their locus in Gaia color--absolute magnitude diagram (CaMD) are also investigated.  More interestingly, for two dwarf novae observed through LAMOST and SDSS in different epoch, their spectra both in quiescence phase and during outburst are exhibited.

\end{abstract}


\section{Introduction} \label{sec:intro}
Cataclysmic variables (CVs) are semi--detached binaries with very short orbital periods, in which white dwarf primaries accrete matter from low--mass secondaries. The secondary are usually late main-sequence stars, and in a few cases, they may also be highly evolved stars such as AM CVn type stars. CVs have been classified according to their amplitudes and timescale of variability, which are classical novae (CN), Recurrent novae (RN), dwarf novae (DN) and nova--like variables (NL). Besides above classification, there is also an important group called magnetic CVs which consist of polars and intermediate polars. Regarding the main characteristics of these objects, a more detailed description can be found in \citet{2003cvs..book.....W}, and \citet{2007ASSL..342.....K}.

Up to now, there are several catalogs of CVs having been published. An early catalog and atlas of CVs was provided by \citet{2001PASP..113..764D} combining and supplementing the first two editions \citep{1993PASP..105..127D, 1997PASP..109..345D}, which finally consists of 1034 CVs containing the basic information of parameters such as coordinates, subtypes magnitudes, and orbital periods etc.. Among these CVs, over 60$\%$ of the objects have available spectra, including 49$\%$ spectra in quiescence and 15$\%$ spectra during outburst. One of the most comprehensive investigations of CV spectra from large--scale sky surveys was performed by \citet{2002AJ....123..430S}. A series of papers (Papers I--VIII) have been published between year 2002 and year 2011 to explore the properties of 285 CVs identified from the Sloan Digital Sky Survey (SDSS) I/II \citep{2002AJ....123..430S, 2003AJ....126.1499S, 2004AJ....128.1882S, 2005AJ....129.2386S, 2006AJ....131..973S, 2007AJ....134..185S, 2009AJ....137.4011S, 2011AJ....142..181S}. Owing to the capability of SDSS to observe faint CVs, \citet{2011AJ....142..181S} analysed the distribution of orbital periods of CVs from SDSS, which is quite different from previous results. Also dedicated to the study of orbital periods of faint CVs from SDSS, \citet{2009MNRAS.397.2170G} discussed the period distribution of a sample which contains a large fraction of short--period CVs. Besides, with the operation of the Catalina Real-time Transient Survey (CRTS), plenty of studies have sprung up to investigate the properties of CVs. \citet{2014MNRAS.441.1186D} provided a sample of 855 CV candidates observed by CRTS, of which 205 targets were discovered for the first time and 137 targets were identified through spectra. \citet{2013yCat..74212414W} presented high--speed photometric observations of dozens of faint CVs selected from SDSS and CRTS. Likewise for faint CVs, \citet{2014MNRAS.443.3174B} examined spectroscopic properties of 85 CVs using both SDSS and follow--up spectroscopic observations, and analysed the photometric properties of 1043 CV systems from CRTS. Moreover,\citet{2015AcA....65..313M} gave the largest sample of 1091 dwarf novae (one subclass of CV) located in Galactic bulge from the Optical Gravitational Lensing Experiment (OGLE) survey. Thus far, large--scale sky surveys, either spectroscopy or photometry, show great potential to search for candidates of CVs.

The Large Sky Area Multi-Object Fiber Spectroscopic Telescope (LAMOST) took its first light in 2008\citep{2012RAA....12.1197C}. After two years of commissioning and one year of pilot survey, LAMOST started a five-year regular survey. The survey mainly focused on the Milky Way and archived several millions of stellar spectra in the Fifth Data Release (DR5).  This massive database is a rich source for detecting CVs, which arouses great interest of researchers being devoted to the studies of CVs. \citet{2013MNRAS.430..986J} proposed a data-mining method by combining the support-vector machine (SVM) with the principal component analysis (PCA) to search for CVs with the commissioning data of LAMOST. They identified a sample of 10 CVs, two of which are new discoveries. \citet{2018RAA....18...68H} collected the spectra of 48 known CVs by cross--matching the published catalogs with LAMOST DR3, and found three new CVs using the method proposed by \citet{2013MNRAS.430..986J}. They investigated not only spectroscopic properties of all these CVs, but also light curves of five CVs through follow--up photometric observations. Common to these CV search efforts is the use of strong emission lines in spectra or some of the properties of known CVs recorded in the literature. So far, there has not been a systematic and comprehensive investigation on CVs using LAMOST data. In this paper, we focus on the search of CVs in different subtypes as well as the analysis of spectral features of each subtype. Further study on orbital periods has begun and the results will be included in a later paper.

We conduct a comprehensive search of CVs in LAMOST DR5 using machine learning methods. According to different spectroscopic characteristics of CV subtypes, we categorize the CVs found in LAMOST DR5 as many as possible. The paper is organized as follows. The observation and the data in LAMOST DR5 are briefly introduced, and the selection process for the initial sample among which we search for CV candidates is presented in section \ref{sec:observations and data reduction}. The methods adopted to search for two groups of CVs with different spectral characteristics are given in detail in section \ref{sec:methods}. Then the sample of CVs from LAMOST DR5 is studied in section \ref{sec:results}, including features and classification of CV subtypes, the distributions in space, in magnitude and in the Gaia CaMD diagram, as well as spectra of common targets both in LAMOST and SDSS datasets. An analysis of 58 new discoveries and a description of the CV catalog are also included in section \ref{sec:results}. Finally, some discussions and summaries for the work are given in section \ref{sec:summary}.

\section{Observations and Data Reduction} \label{sec:observations and data reduction}
Undertaken by the Chinese Academy of Science, the LAMOST survey has become the first spectroscopic survey to collect tens of millions spectra from the universe. Owing to the unique design of LAMOST, it is able to acquire 4000 spectra simultaneously in a single exposure with a limiting magnitude as faint as r$\sim$18 at the resolution of 1800. The wavelength range for these low resolution spectra covers the optical band from 3700$\AA$ to 9000$\AA$. The LAMOST survey contains two major components including the LAMOST ExtraGAlactic Survey (LEGAS), and the LAMOST Experiment for Galactic Understanding and Exploration (LEGUE) survey \citep{2012RAA....12..723Z, 2012RAA....12..735D}. The raw data has been processed by LAMOST pipelines to extract, calibrate as well as classify the spectra, which were described in detail in \citet{2012RAA....12.1243L, 2015RAA....15.1095L}. With the released spectra,  objects have been classified as galaxies, QSOs, and stars, and stars have been further classified into seven subtype along the temperature sequence as well as special type such as white dwarfs, double stars, carbon stars etc.

So far, the LAMOST survey has been conducted for about 8 years since 2011, including a one-year pilot survey followed by an on--going regular survey. The LAMOST--I regular survey was defined as from September 2012 to June 2017, which covers approximately 17,000 square degrees of the sky. The data of LAMOST--I survey together with the pilot survey make up the fifth data release (DR5) which has been available to the public in the June of 2019 \citep{2019luoal}. The total number of spectra in DR5 is more than 9 million, including 152,863 galaxies, 52,453 QSOs, 8,183,160 stars and 637,889 objects with unclassified type. Among the massive data, there are about 6 million spectra of which the signal--to--noise ratios (SNR) per pixel for both g band and i band are larger than 10. Along with the general catalog, another three specific samples selected from the whole sample are also released as separate catalogs, including a catalog of 5,348,712 late--A, F, G and K stars with the atmospheric parameters (i.e.T$\emph{eff}$, logg and [Fe/H]), a catalog of $\sim$ 439,920 A--type stars and a catalog of $\sim$ 534,393 M dwarfs.

In this work, we carry out a comprehensive search of CVs in LAMOST DR5, and we first need to define a search range in the whole released dataset. Considering the spectral characteristics, a large portion of CVs were probably mis-classified as O/B/A stars or white dwarfs by LAMOST pipeline because of the template matching algorithm it used. Even for those having been classified as CVs by the pipeline, we find that over 97$\%$ of them are actually spectra of normal stars which are contaminated by superimposing some nebular emission lines from HII regions, and only $\sim$ 3$\%$ are spectra of real CVs. To collect as much CV spectra as possible, the ``Unknown” objects in DR5 should also be taken into account since the pipeline refuses to classify some ``bad" spectra which results in a few CVs hidden in the ``Unknown” dataset. To sum up, we select a total of $\sim$ 1,130,000 spectra to be the initial sample we begin with the search, including $\sim$ 480,000 O/B/A stars, $\sim$ 20,000 white dwarfs, $\sim$ 3,000 CVs and $\sim$ 630,000 ``Unknown" spectra.

\section{Methods} \label{sec:methods}
According to the spectral characteristics, we search for two types of CV spectra which display significantly different features. One type of CV spectra are dominated by obvious emission lines, which are probably Dwarf Nova in quiescence or Nova--like variables. The other type of CV spectra show broad absorption lines where emission lines are overwhelmed by their continuum, which are probably CVs in outburst status or CVs surrounding with optically thick disks. We need to adopt different methods to select the candidates of these two groups respectively.

For both two groups with specific features of CVs, some known CV spectra are needed as templates to pick up the candidates in the initial dataset. We first collect the CVs catalogues from the literatures as well as the SIMBAD database, which eventually contain 4,215 CVs without being removed duplicate sources \citep{1993PASP..105..127D, 2000A&AS..143....9W, 2001PASP..113..764D, 2011AJ....142..181S, 2014MNRAS.437..510C, 2014MNRAS.441.1186D, 2014MNRAS.443.3174B, 2016MNRAS.456.4441C}. Cross-matching known CVs and LAMOST catalogs within a cross radius of 5$\arcsec$, about 500 LAMOST spectra of known CVs are obtained. After visually checking these spectra, more than half of the spectra are difficult to be recognized as CVs due to their low SNR ( $<$ 3). Besides, There are also a small number of spectra showing general characteristics of F, G, or K--type spectra without any emission lines, which are likely not CVs. Finally, 155 spectra are identified to display the detectable characteristics of CVs. Among the 155 CV spectra, we select 75 high--quality spectra as templates to search for two groups of CVs separately, in which 50 spectra are in one group having strong emission lines and 25 spectra are in another group with broad absorption lines.

\subsection{Spectra Dominated by Emission Lines}
The CV spectra of this group are characterized by their significant emission lines, such as Balmer series and/or He I/He II. Such obvious features could be easily hunted in a big dataset using a so called "BaggingTopPush" algorithm developed by \citet{2016PASP..128c4502D}, which combines a bipartite ranking model with the bootstrap aggregating techniques. The core idea of the BaggingTopPush approach is to build a ranking model that ranks the unlabelled samples based on their relevance scores to the positive samples (the given templates). The method was effectively applied to search for carbon stars from SDSS DR10 and LAMOST DR4 performed by \citet{2016PASP..128c4502D, 2018ApJS..234...31L}. A detailed introduction and application of this method can be referred in \citet{2016PASP..128c4502D}. The reason why we adopt this method is considering its advantage that the expert knowledge can be taken into account when checking the ranked result.

\begin{figure}[ht!]
\plotone{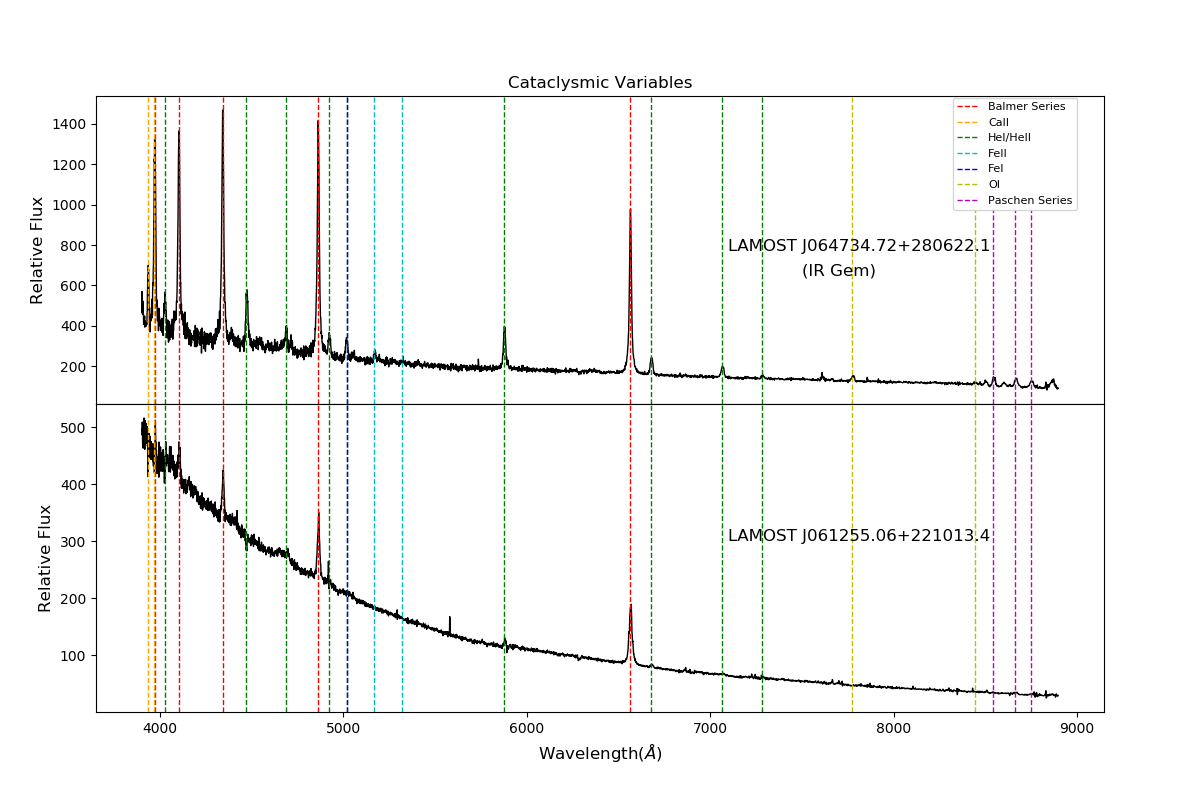}
\caption{Two examples of CV spectra with significant emission lines from LAMOST survey. Some strong emission lines are marked in different colors in the spectra. We also shown the LAMOST designations of objects in figures. \label{Figure 1}}
\end{figure}

In the training stage, 50 CV templates collected from the known CV catalogues mentioned above are considered as the positive samples. Figure \ref{Figure 1} shows two examples of CV spectra as templates from LAMOST DR5. All the spectra of this type display strong emissions of Balmer series including  H$\alpha$, H$\beta$,  H$\gamma$, H$\delta$ and H$\epsilon$. Other emission lines such as He I, He II and Fe II also frequently appear in these spectra, not as strong as Balmer emission lines. According to the spectral features, a CV spectrum dominated by strong emission lines is probably classified as of either an O/B/A--type star, a white dwarf, a CV or even an ··UNKNOWN“  object by LAMOST pipeline. Thus we need to search for this kind of CV in the whole initial dataset containing 1,130,000 spectra, which is divided into ten groups to apply the BaggingTopPush. Empirically, we manually check the result of top one thousand spectra for each of the ten groups. We notice that the false CVs in the top thousand spectra are mainly spectra superimposed nebula emissions. In addition, the lower ranking, the fewer CVs are found, and there are almost no CVs in the bottom 500 objects. We eventually obtain 256 spectra of CV candidates exhibiting strong emission lines. 

\subsection{Spectra with Broad Absorption Lines}
\begin{figure}[ht!]
\plotone{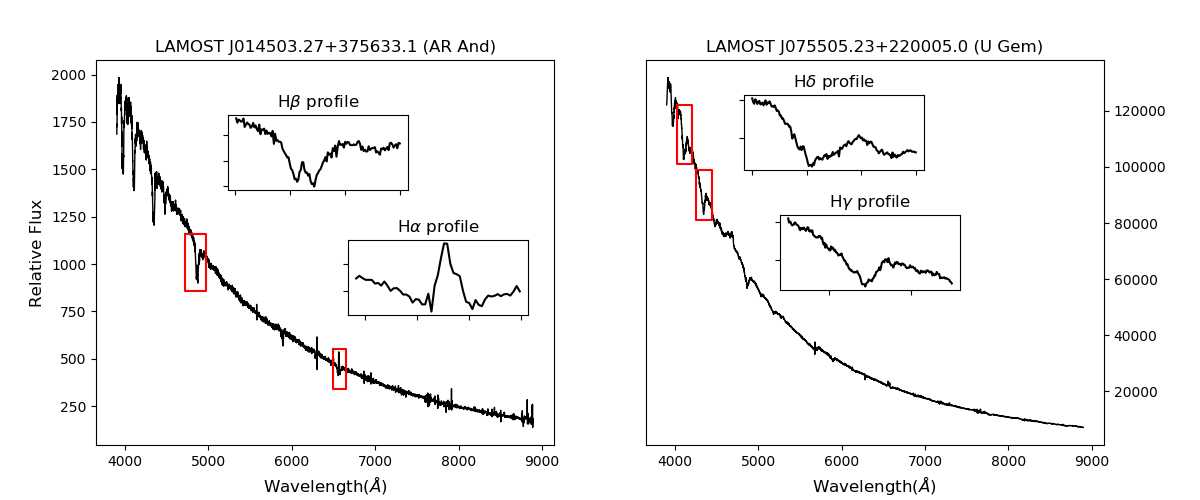}
\caption{Two spectra of CVs with broad absorption lines from LAMOST survey, which show identical spectral characteritics. The profiles of H$\alpha$, H$\beta$ of the left spectra and H$\gamma$ and H$\delta$ of the right spectra are plotted in the figure. Emission components appear in both H$\alpha$ and H$\beta$ lines, but are absent in H$\gamma$ and H$\delta$. \label{Figure 2}}
\end{figure}

Another group of CV spectra exhibit their main feature of broad absorption spectral lines along with weak emissions in the core (mostly in H$\alpha$ or H$\beta$). As mentioned above, these objects contain different types of CVs, such as dwarf nave in outburst and CVs with lower disk luminosity. An example of two spectra of this type from LAMOST survey are shown in Figure \ref{Figure 2}, which show the bright blue continua and broad absorption Balmer lines. It is likely that the LAMOST pipeline misclassified nearly all of this kind of CV as of O/B/A stars or white dwarfs due to the high similarity between O/B/A stars or white dwarfs with CVs in spectra. Therefore, we select this kind of CV spectra from O/B/A stars, white dwarfs and CVs in DR5 which include more than 470,000 spectra. In addition, since spectral features of this kind of CV are not as prominent as CV spectra with emission lines, we use both the BaggingTopPush approach and the method of Random Forest based on the width and depth of Balmer lines to select CV candidates of this type as many as possible.

\subsubsection{Preliminary Selection}
In the spectra of these CVs, low--order Balmer lines such as H$\alpha$ and H$\beta$ always show emissions in the cores, and higher order of Balmer series show broad absorption lines. Therefore, we firstly pick up spectra with H$\alpha$ in emission from more than 470,000 O/B/A, white dwarf and CV spectra using the method proposed by \citet{2016RAA....16..138H}. 
The method aims to figure out two different profile shapes of strong H$\alpha$ emission lines as well as weak emission lines lying in the deep H$\alpha$ absorption profiles in spectra. By applying the criteria listed in the following, we select about 17 thousand spectra with H$\alpha$ emission lines from $\sim$ 470,000 spectra. 

\begin{equation} \label{detection1}
 \sum_{i=-5}^{5}f_{obs}[n_{0}~+~i]~/~11 ~>~ f_{conti}[n_{0}]
\end{equation}

\begin{equation} \label{detection2}
 \begin{aligned}
 & \sum_{i=-1}^{1}f[n_{0}~+~i]~/~3 > \sum_{i=-2}^{2}f[n_{0}~+~i]~/~5 ~~~~~~~\& \\
 & max(f_{obs}[n_{0}-1:n_{0}+1]) \geqslant max(f_{obs}[n_{0}-2:n_{0}+2]) 
 \end{aligned}
\end{equation}
where n$_{0}$ is the pixel index of central wavelength of H$\alpha$ ($\lambda_{0}$ = 6564 \AA), $f_{obs}$, $f_{conti}$ denote the flux for observed spectra and continuum spectra respectively, and max(f$_{obs}$[x:y]) represents the maximum flux where the index is from x to y.

\subsubsection{the BaggingTopPush Approach}
The same as the method used in section 3.1, we also adopt the BaggingTopPush approach to select CVs with broad absorption lines from about 17 thousand spectra showing H$\alpha$ emissions. 25 CV templates with broad absorption lines picked up from the known CVs are considered as the positive samples when we apply the method to 17 thousand spectra. Similarly, we checked the top ten thousand spectra by experience and finally pick up 91 CVs showing broad absorption lines. However, due to the less significant characteristics of this kind of CV spectra, it is not as effective as searching for CV spectra with strong emission lines in section 3.1. A more efficient classification method are adopted as an important supplement for picking up CV spectra with broad absorption lines, which is described in the following subsection.

\subsubsection{Spectral line based Random Forest}
Two obvious differences between the spectra of CVs with broad absorption lines and another type with H$\alpha$ emissions are the line widths and depths of Balmer series. Different with 17 thousand spectra with H$\alpha$ emissions,  H$\alpha$ and H$\beta$ of these CV spectra often show emission cores, while higher order of Balmer series such as H$\gamma$ and H$\delta$ basically have pure broad absorption lines. So we fit H$\gamma$ and H$\delta$ lines by gaussian function to obtain both the line widths and line depths, which can be used to distinguish CVs from other type stars. We then use the Random Forest (RF) algorithm to effectively select samples of CVs based on these line widths and depths. 

\begin{figure}[ht!]
\plotone{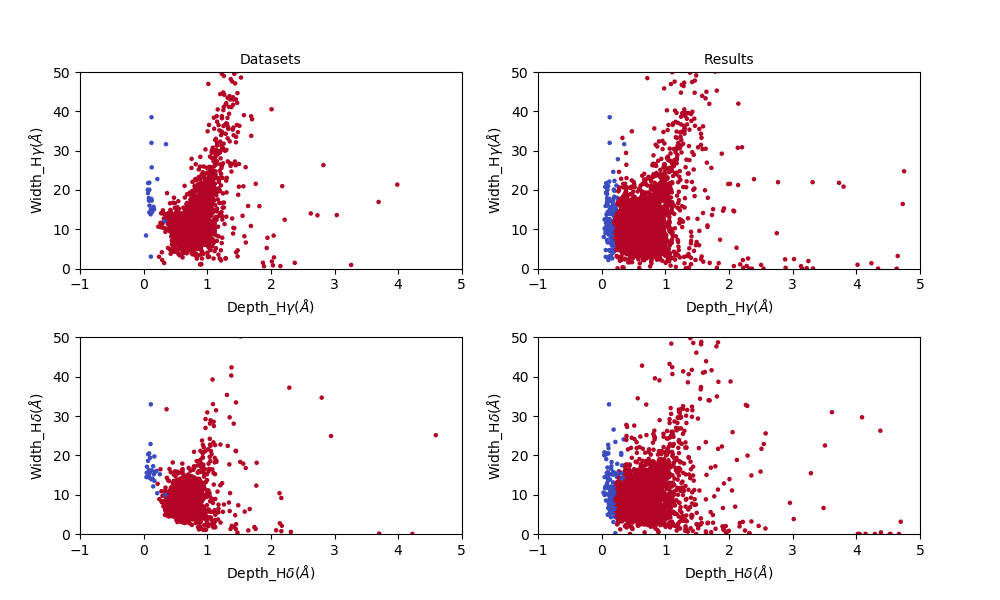}
\caption{The result of CV candidates using Random Forest algrithm based on H$\gamma$ and H$\delta$ line features. The positive and negative datasets are shown in the left two panels and the result using the method of Random Forest are shown in the right panels. The CV spectra are represented by blue points and O/B/A--type spectra are plotted as red points. \label{Figure 3}}
\end{figure}

The 25 absorption-line CVs mentioned above are fed to the RF as positive templates. For the negative dataset, we randomly select 5000 spectra classified by the LAMOST pipeline as normal O/B/A-type. As shown in Figure \ref{Figure 3}, the left two panels show the distributions of line features of templates, in which the CV templates are plotted in blue and the O/B/A--type spectra are in red. The line features of H$\gamma$ and H$\delta$ of $\sim$170,000 spectra from LAMOST DR5 are plotted by red points in the right panels, and the blue ones represent the CV candidates selected by the Random Forest using the line features of H$\gamma$ and H$\delta$. The final CV candidates are the intersection dataset of RF results from H$\gamma$ line feature and H$\delta$ line features, and they are also checked by the careful manual inspection. In combination with the results of the BaggingTopPush approach and the RF method, we finally pick out 124 CV candidates with broad absorption lines.

\section{Results} \label{sec:results}
A total of 380 spectra of CVs which correspond to 245 unique objects are found in LAMOST DR5. By cross--matching with 4,976 known CVs collected from the previous literatures, 58 out of 245 are newly discovered candidates that have no record in the published CV catalogs.

\subsection{Spectral Features}
For the 380 CV spectra, we calculate the equivalent widths of prominent Balmer lines, HeI $\lambda$4471 and HeII $\lambda$4686, as well as the ratio of H$\beta$ to HeII $\lambda$4686. For most CVs with strong emission lines, the Balmer decrement from H$\alpha$ to H$\delta$ becomes shallow or sometimes even negative which is consistent with \citet{2003cvs..book.....W}. 

Although the classification of CVs is generally based on their amplitudes and timescale of the variabilities, we can still see some clues through spectral features. The main spectral characteristics to classify CV subtypes include HeII emission lines, FeII emission lines, high--excitation emission lines such as CIII and NIII, as well as the line ratio of  H$\beta$ to HeII $\lambda$4686. In order to determine the subtype of CVs, we manually inspect all the spectra, compare the intensities of H$\beta$ with HeII $\lambda$4686, and check the existence of unique spectral lines in each subtype spectra. Based on the spectral features, we assign subtypes to as many of the CVs as possible. A detailed description of the classification follows.

\begin{figure}[ht!]
\plotone{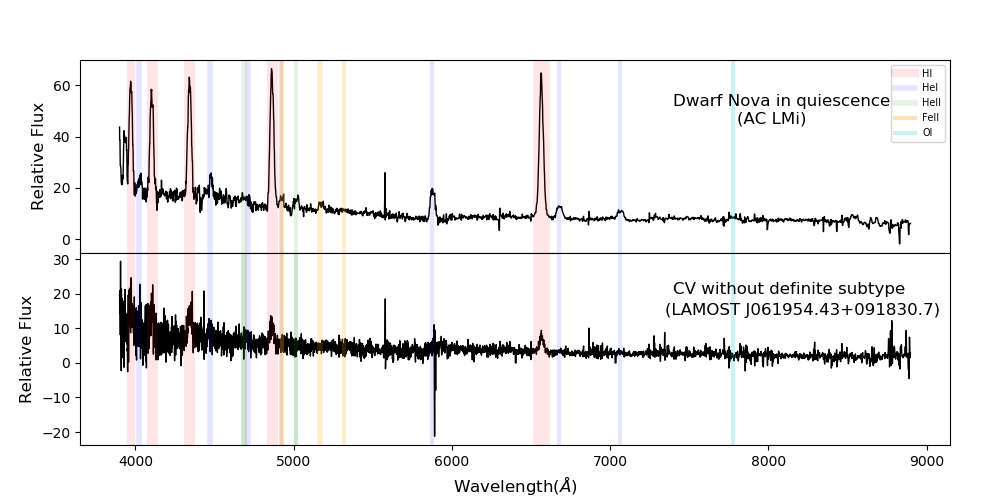}
\caption{The spectra of DNs in quiescence and CV without a definite subtype from LAMOST DR5. The common spectral lines shown in DN spectrum are marked by different colors for different elements, including HI, HeI, HeII, FeII and OI. A spectrum of CV without a definite subtypes is plotted in the bottom panel. Due to its low SNR, the spectral features can't be seen clearly except strong Balmer emission lines.\label{Figure 4}}
\end{figure}

\textit{Dwarf Nova (DN):} Table \ref{tab1:DNline} shows the basic emission lines appearing in the spectra of DNs in quiescence. For spectra of quiescent DNs, prime characteristics include large ratio of H$\beta$ to HeII $\lambda$4686 as well as a weak or absent blended CIII/NIII $\lambda$4650 \citep{2003cvs..book.....W}. We check 95 quiescent spectra in our sample which are classified as DN in the previous literatures. As a result, the value of H$\beta$/HeII $\lambda$4686 are larger than 2 for almost all the spectra of DNs in quiescence, except those spectra that have too low SNR to calculate the accurate line intensity. Another important characteristic is that Fe emission lines always appear in the spectra of DNs, such as FeII $\lambda$4924, $\lambda$5169 and $\lambda$5317. According to the criteria mentioned above, CV spectra from LAMOST DR5 which have the value of H$\beta$/HeII $\lambda$4686 $>$ 2 and display the emission line of iron such as FeII $\lambda$5169 are regarded as spectra of quiescent DNs shown in Figure \ref{Figure 4}. The significant spectral lines in spectra of quiescent DNs are marked by different colors in Figure \ref{Figure 4}. The FeII emission lines selected as one criterion of DN are marked by yellow in the figure, which include FeII $\lambda$5169, $\lambda$5317 and FeII $\lambda$4924 blending with HeI $\lambda$4922. There are also a portion of CV spectra showing no obvious emission lines besides Balmer series, like the spectrum in the bottom panel of Figure \ref{Figure 4}. Because most of the spectral characteristics are overwhelmed by noise, we are unable to make a definite subtype classification.

\begin{figure}[ht!]
\plotone{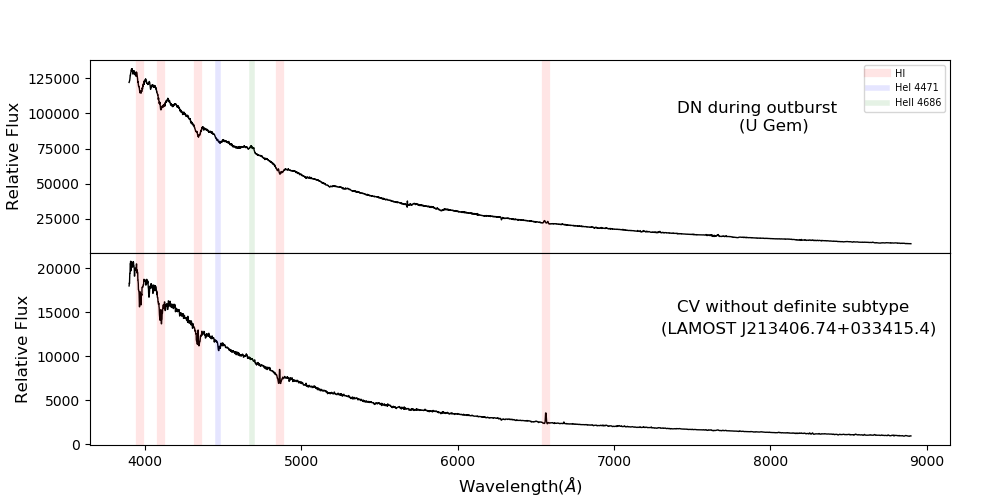}
\caption{The spectra of DNs undergoing outburst and CV without a definite subtype from LAMOST DR5. The main absorption features and wide emission line HeII $\lambda$4686 shown in the DN during outburst are marked by different colors for different elements in the top panel. For the CV in the bottom panel, broad absorption lines such as Balmer series and HeI $\lambda$4471 are also shown in the spectrum. Besides, there are no obvious emission lines, apart from H$\alpha$ emission and narrow emission cores in higher order of Balmer line profiles. \label{Figure 5}}
\end{figure}

In addition, we also find several spectra of DNs undergoing outburst. For these spectra, the distinct spectral characteristics include the absorption lines such as Balmer series and HeI $\lambda$4471, which have the similar widths to those of emission lines in quiescent spectra as well as a wide HeII $\lambda$4686 emission line. A spectrum of DN undergoing outburst is shown in the top panel of Figure \ref{Figure 5}, in which main broad absorption lines and the HeII $\lambda$4686 emission line are marked by different colors. Besides, some other spectra also show similar absorption features to those of DN undergoing outburst, as seen in the bottom panel of Figure \ref{Figure 5}. In these spectra, narrow emission cores appear within the broad absorption lines. It indicates either DNs on the decline from outburst, or CV systems with low disk contribution where underlying stars provide most of the luminosity. Therefore, we don't provide the definite subtypes for these CV spectra. Finally, we classify 134 spectra in our sample as DN candidates.

\begin{figure}[ht!]
\plotone{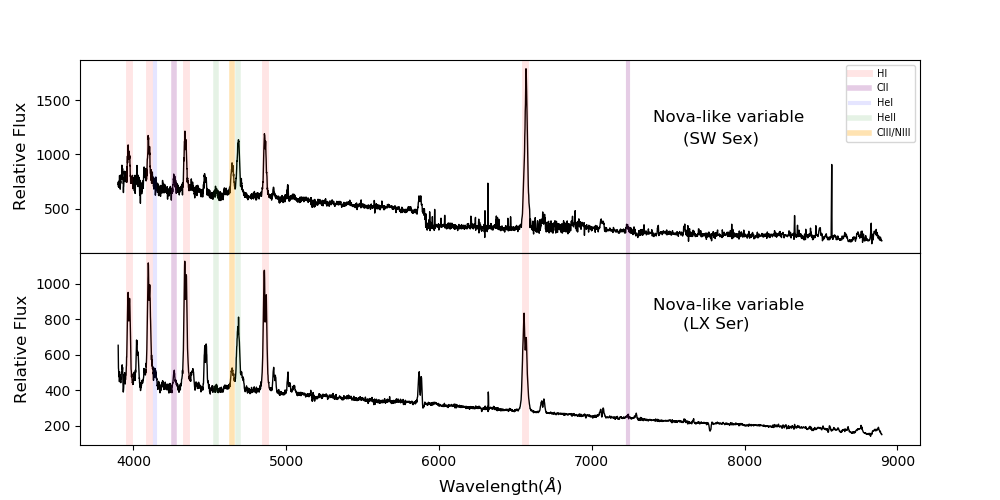}
\caption{The spectra of NL from LAMOST DR5. Balmer series, wide emission line HeII $\lambda$4686 and CIII/NIII $\lambda$4650 blend are shown in the spectra. In addition, some unique spectral lines in NL spectra including CII and HeI are also shown. These emission lines are marked by different colors for different elements. \label{Figure 6}}
\end{figure}

\textit{Nova--like Variables (NL):} Most of the spectral lines listed in Table \ref{tab1:DNline}, such as HeI, appear in the spectra of NL. Besides, there are also some other emission lines, especially more and stronger high excitation emission lines present in NL spectra which are listed in Table \ref{tab2:NLline}. One critical characteristic is that HeII $\lambda$4686 and/or  CIII/NIII $\lambda$4650 are relatively stronger, and the former line may even be stronger than H$\beta$ in NL spectra. Moreover, a few additional spectral lines (as seen in Table \ref{tab2:NLline}) are also often detectable in NL spectra, such as CII $\lambda$4267, $\lambda$7234 and HeII $\lambda$4542. Similarly, we examine the ratio of HeII $\lambda$4686 to H$\beta$ for spectra which are marked as NL in literatures, and find that 14 out of 18 spectra of NL have the values larger than 0.5. Therefore, CVs which have strong HeII $\lambda$4686, H$\beta$ emission lines, and the value of HeII $\lambda$4686/H$\beta$ $>$0.5 are considered as the candidates of NL. In addition, an obvious emission of CIII/NIII $\lambda$4650 appearing in the spectra is also considered as the criterion of NL candidates. Meanwhile, We make a further identification by individually checking whether the spectra have the high excitation emission lines such as CII $\lambda$4267, $\lambda$7234. As a result, 41 spectra in our sample are classified as NL candidates. Spectra of NL from LAMOST DR5 are shown in Figure \ref{Figure 6}. Apart from Balmer series, HeII $\lambda$4686 and CIII/NIII $\lambda$4650 blend, we also mark some unique emission line in NL spectra by different colors.

\begin{figure}[ht!]
\plotone{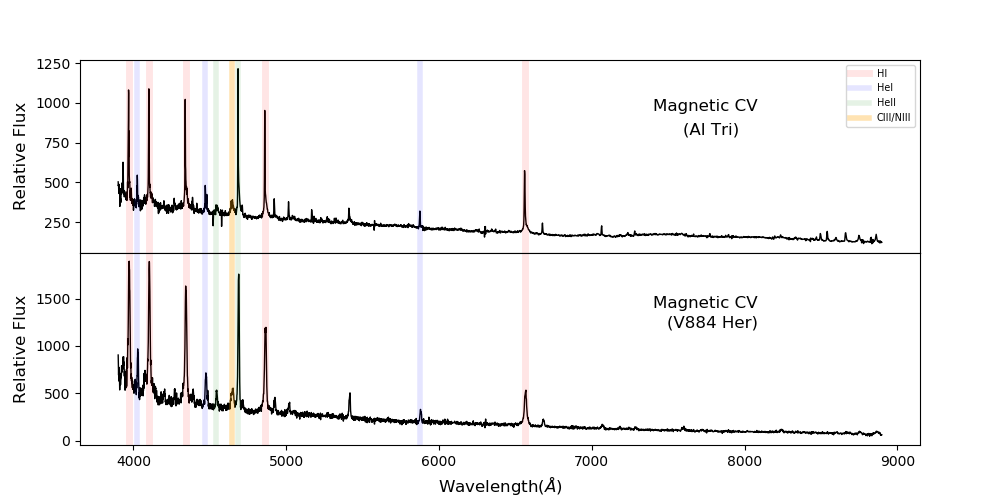}
\caption{The spectra of magnetic CVs from LAMOST DR5. The significant spectral lines including Balmer series, HeII $\lambda$4686 and CIII/NIII $\lambda$4650 blend are shown in the spectra. These emission lines are marked by different colors for different elements. \label{Figure 7}}
\end{figure}

\textit{Magnetic CVs:} According to \citet{2003cvs..book.....W}, magnetic CVs are usually included among NL, and the polars and the intermediate polars are two types of magnetic CVs. From the perspective of spectral classification, the spectra of this group of CVs are distinguished using the prominent characteristic lines of HeI, HeII and CIII/NIII $\lambda$4650 blend.  Additionally, the strength ratio H$\beta$ / HeII $\lambda$4686  could be used to distinguish polars and intermediate polars. For polars, these two lines are comparable, while for intermediate polars, HeII $\lambda$4686 is generally slightly weaker than H$\beta$\citep{2003cvs..book.....W}. However, because the characteristics mentioned above resemble those of NL, we are not able to provide an accurate classification. Therefore, CVs which have the ratio of HeII $\lambda$4686 to H$\beta$ larger than 1 as well as strong HeI, HeII and CIII/NIII blend at 4650\AA  ~are considered as the candidates of magnetic CVs. Among the sample, 19 spectra are classified as the candidates of magnetic CVs. Two spectra of magnetic CV from LAMOST DR5 are shown in Figure \ref{Figure 7}. As seen in Figure \ref{Figure 7}, HeII $\lambda$4686 is extremely strong in both spectra, which intensity obviously exceeds that of H$\beta$ indicating a probable existence of magnetic field.

\begin{table}[h!]
\centering
\caption{A basic emission--line list for spectra of quiescent DN in the optical region \label{tab1:DNline}}
\begin{tabular}{clclcl}
\tablewidth{0pt}
\hline
\hline
Wavelength($\AA$) & Spectral Line & Wavelength($\AA$) & Spectral Line & Wavelength($\AA$) & Spectral Line \\
\hline
\decimals
3934 & CaII & 4861 & H$\beta$  & 8438 & P18  \\
3968 & CaII & 4922 & HeI  & 8446 & OI  \\
3970 & H$\epsilon$ & 4924 & FeII  & 8467 & P17  \\
4026 & HeI & 5016 & HeII  & 8498 & CaII  \\
4102 & H$\delta$ & 5018 & FeI  & 8502 & P16  \\
4121 & HeI & 5169 & FeII  & 8542 & CaII  \\
4634-42 & NIII & 5317 & FeII  & 8545 & P15  \\
4340 & H$\gamma$ & 5876 & HeII  & 8598 & P14  \\
4388 & HeI & 6563 & H$\alpha$  & 8662 & CaII  \\
4471 & HeI & 6678 & HeI  & 8665 & P13  \\
4634-42 & CIII & 7065 & HeI  & 8750 & P12  \\
4647-51 & NIII & 7281 & HeI  & 8863 & P11  \\
4686 & HeII & 7772,74,75 & OI  & 9015 & P10  \\
4713 & HeI & 8236 & HeII  & 9229 & P9(+FeII,MgII) \\
\hline
\end{tabular}
\end{table}

\begin{table}[h!]
\centering
\caption{Additional emission lines in the spectra NL variables in the optical region.\label{tab2:NLline}}
\begin{tabular}{clcl}
\tablewidth{0pt}
\hline
\hline
Wavelength($\AA$) & Spectral Line & Wavelength($\AA$) & Spectral Line \\
\hline
\decimals
3819 & HeI & 4267 & CII  \\
4070-76 & OII & 4415-17 & OII  \\
4089 & SiII & 4542 & HeII  \\
4128-30 & SiII & 5805 & CIV  \\
4143 & HeI & 7234 & CII  \\
\hline
\end{tabular}
\end{table}

\begin{figure}
\plotone{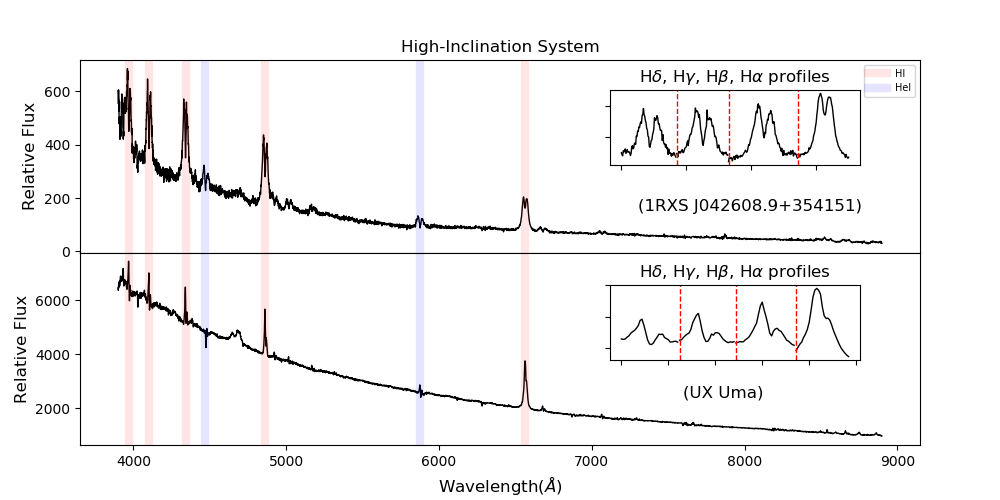}
\caption{Spectra of high--inclination systems from LAMOST DR5. Spectral line profiles exhibiting double peak can be seen clearly in the spectra, such as Bamer series and HeI emission lines. Double--peak profiles are marked by different colors for different elements. Besides, a close-up of H$\alpha$, H$\beta$, H$\gamma$ and H$\delta$ profiles for each spectrum is shown in the subgraph.\label{Figure 8}}
\end{figure}

\textit{High-Inclination System:} The systems with high inclination typically show very prominent central absorption in Balmer emission lines with the central depression sometimes reaching to the continuum level around the higher order lines. 89 CVs in our sample showing obvious double--peaked emission lines are probably high--inclination systems. Examples of possible high--inclination systems are shown in Figure \ref{Figure 8}, in which the Balmer line profiles for both spectra are also zoomed in.

\begin{figure}
\plotone{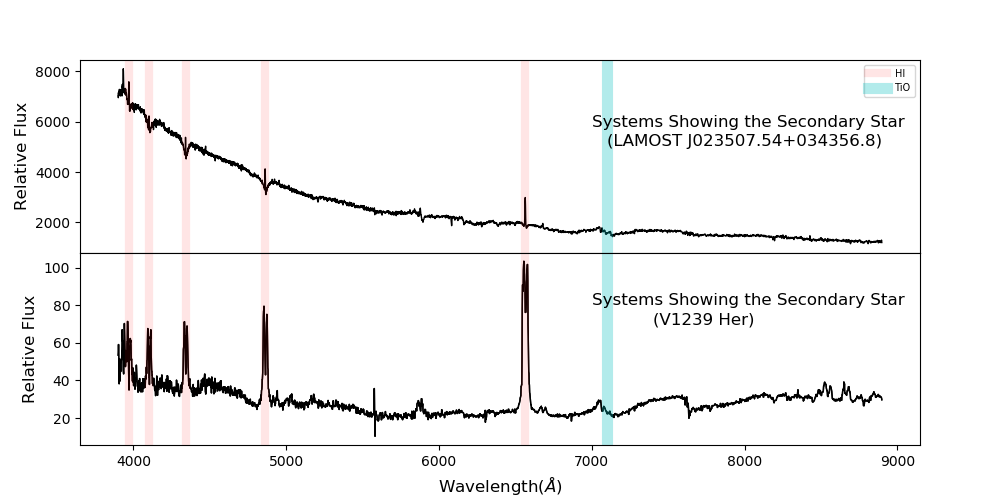}
\caption{Spectra of CV systems showing secondary stars from LAMOST DR5. Both spectra exhibit the characteristic of M--type star at 7100 $\AA$ band. For the top spectrum, the emission lines are quite narrow lying in the broad aborption lines. For the bottom spectrum, the emission lines show deep central absorption reverses, which is typical feature of a high--inclination system. \label{Figure 9}}
\end{figure}

\textit{Systems Showing the Secondary Stars:} By inspecting 380 spectra of CVs, there are 33 systems that show spectral signatures from the secondary because of the appearance of TiO band at 7100 \AA. This obvious feature reveals that each of them has an M--type secondary, probably indicating the existence of a tenuous disk with a low mass transfer rate, or a pre--CV system without a mass transfer rate, of which the emission lines are generated by the secondary star. There is another explanation that the low disk luminosity may be due to the high inclination effect, such as the system plotted in the bottom panel of Figure \ref{Figure 9}. For all the cases, the primary and secondary stars of the systems contribute most of the luminosity so that spectral features of the underlying stars can be observed. Figure \ref{Figure 10} shows two spectra of CVs showing the M secondaries. In the spectrum of top panel, the feature of 7100 $\AA$ band indicates the presence of an M--type star, and very narrow emission--line cores appear in Balmer series which are likely generate by the M secondary. In the bottom panel, the CV spectrum exhibits not only the characteristics of a companion M star, but also double--peak features of spectral lines. Therefore, the CV probably has an edge--on disk which contribute a small portion of light.

\subsection{Basic Distributions}
\begin{figure}
\plotone{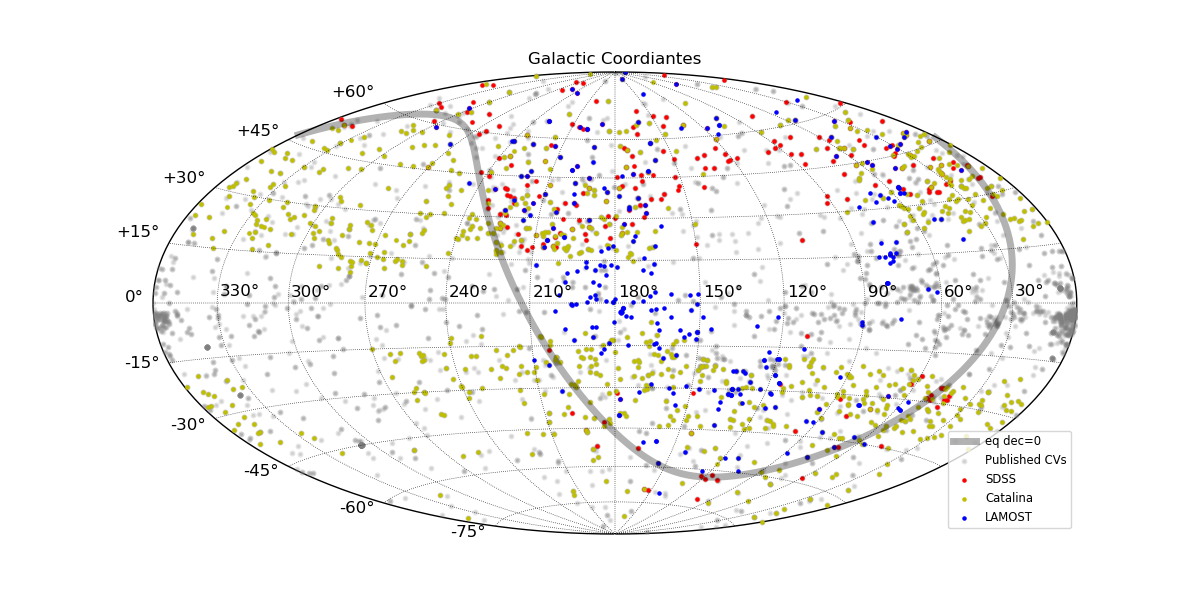}
\plotone{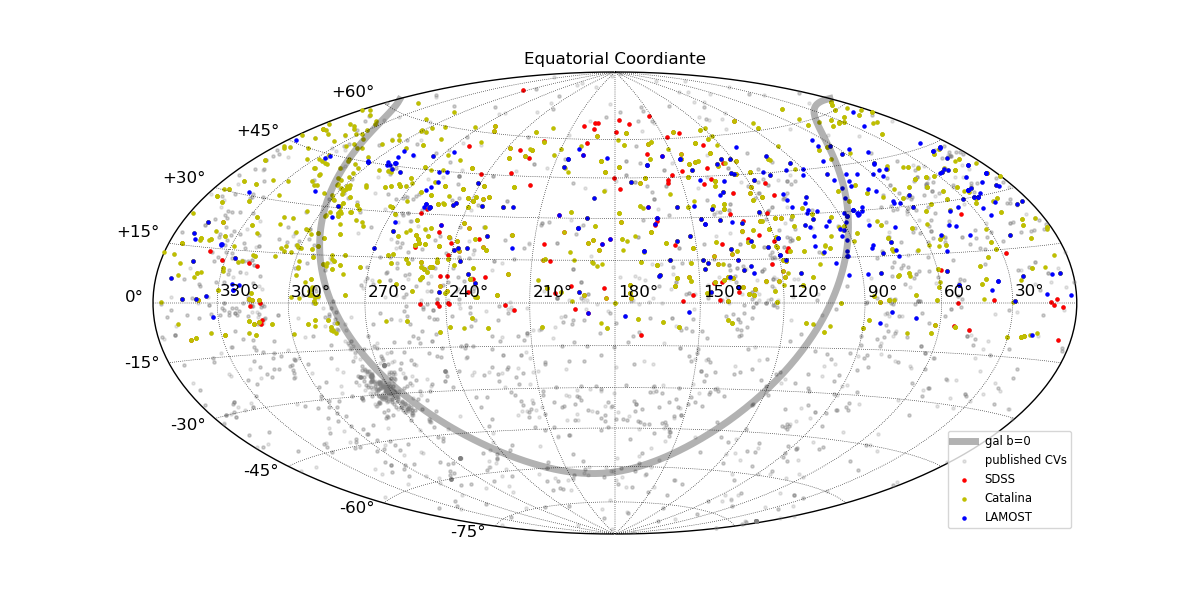}
\caption{Spatial distributions of CVs from different catalogues in galactic and equatorial coordinates. For both coordinates systems, the gray points represent all CVs collected from published catalogs. In particular, CVs from SDSS and Catalina survey are plotted by red and yellow points respectively. CVs selected from LAMOST DR5 are represented by blue points. The spatial distributions of CVs from these three surveys are different from each other. \label{Figure 10}}
\end{figure}

To better understand the CV samples found in LAMOST DR5, we present some distributions of the sample. Figure \ref{Figure 10} shows the spatial distributions of CVs from LAMOST, SDSS, Catalina and the published catalogs. For CVs from SDSS survey, the targets are distributed in the area of higher galactic latitude. For CVs from Catalina survey, there is a gap existing in the galactic plane owing to its observation strategy. While for the sample selected from LAMOST DR5, the spatial distribution of CVs follows the footprint of LAMOST survey, in which part of CVs are found at the low galactic latitude, especially in the direction of the Galactic Anti-center. 

\begin{figure}
\plotone{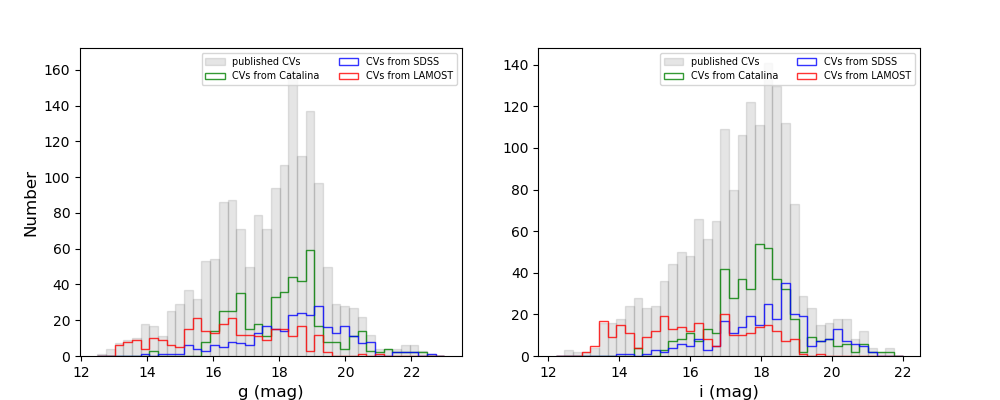}
\caption{The magnitude distributions of g and i band for CVs from published catalogs, Catalina, SDSS and LAMOST survey. \label{Figure 11}}
\end{figure}

The magnitude distributions of g and i bands for CVs from published catalogs, Catalina, SDSS and LAMOST survey are shown in Figure \ref{Figure 11}. From the figure, we can see that CVs from the spectroscopic surveys of SDSS and LAMOST are respectively distributed in the faint and bright ends. In general, CVs from LAMOST data are about 2 or 3 magnitude brighter than those from Catalina and SDSS. 

\begin{figure}
\plotone{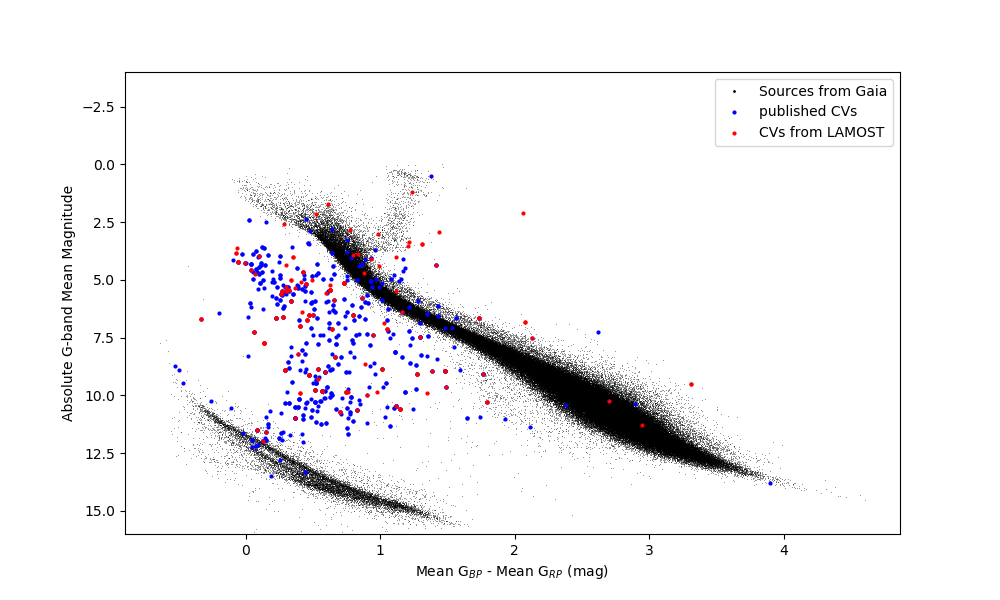}
\caption{CVs from published catalogs and LAMOST in Gaia CMD. \label{Figure 12}}
\end{figure}

We also use the astrometric and photometric data from Gaia to plot CVs in the CaMD. CVs found in this work together with those published in previous catalogs are corss-matched with the Gaia data, their parallax, the mean G--band photometry and the color of mean G$_B$$_P$ - mean G$_R$$_P$ are obtained. To compare the locus of CVs in CaMD with the main sequence and white dwarf sequence, a sample of 300,000 objects are selected from Gaia of which the parallaxs are greater than 10 mas and the relative parallax precisions are better than 20$\%$. Figure \ref{Figure 12} shows the Gaia CaMD, and we can see that the majority of CVs are located between the main sequence (the upper branch across diagonals) and the white dwarf sequence (the lower branch).

\subsection{New Candidates}
In the sample of CVs spectroscopically selected from LAMOST DR5, 71 spectra for 58 objects are new discoveries, which are shown in Figure \ref{Figure_append}. It should be pointed out that some artifacts may appear in the middle of spectra, where the blue arms are combined with the red ones. The combination region in LAMOST spectra is between 5700 and 5900 $\AA$, which sometimes shows unreliable characteristics such as bumps due to the data procession. For example, a drop in the middle of spectrum of J011726.67-083026.9 and a steep rise in that of J052254.86+402352.3 are both caused by the processing method of combining blue with red arms. Another examples are J052201.61+373806.0, J053448.33-063026.8 and J053036.55+341950.5, which have broad emission lines in the combination region of the spectra. These characteristics are also very likely brought by the processing method, which must be cautiously dealt with when taking them into account. The distributions of SNR for CVs and new candidates selected from DR5 are shown in Figure \ref{Figure 13}. Among these objects, 31 spectra show blue continua and broad absorption Balmer lines, which are probably the DN undergoing outburst, NL similar to UX UMa, or any of the other systems with optically thick disks. Strong He II 4686 emission lines appear in the 31 spectra, in some of which the strength of He II are comparable to that of H$\beta$. Especially, there are eight spectra showing the likely polar nature that He II emission lines are much stronger than H$\beta$ and the broad CIII/NIII 4650 $\AA$ blends are also present. Further identification needs more follow--up observations such as using the light curves to study their amplitudes and time scales of the variations. Besides, 12 spectra present absorption profiles with very narrow emission lines, among which there are eight spectra also showing the feature of M--type secondary. These new discoveries might belong to the group of pre--CVs.

\begin{figure*}
\plotone{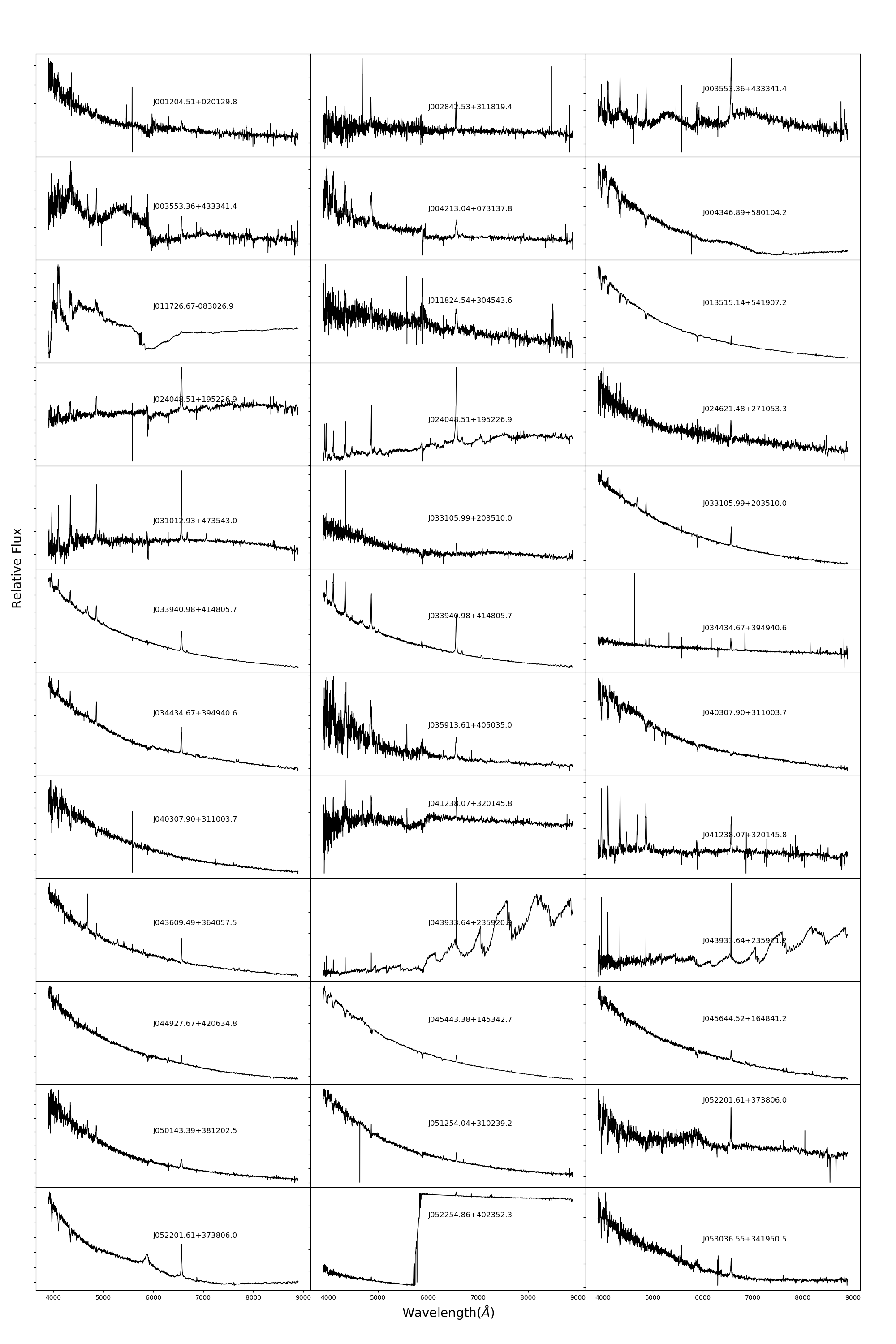}
\caption{71 spectra of 58 new discoveries. The horizontal X--axis represents the wavelength coverage ranging from 3900 to 8900 $\AA$, and the vertical Y--axis represents the relative flux. \label{Figure_append}}
\end{figure*}

\addtocounter{figure}{-1}
\begin{figure*}
\addtocounter{figure}{0}
\plotone{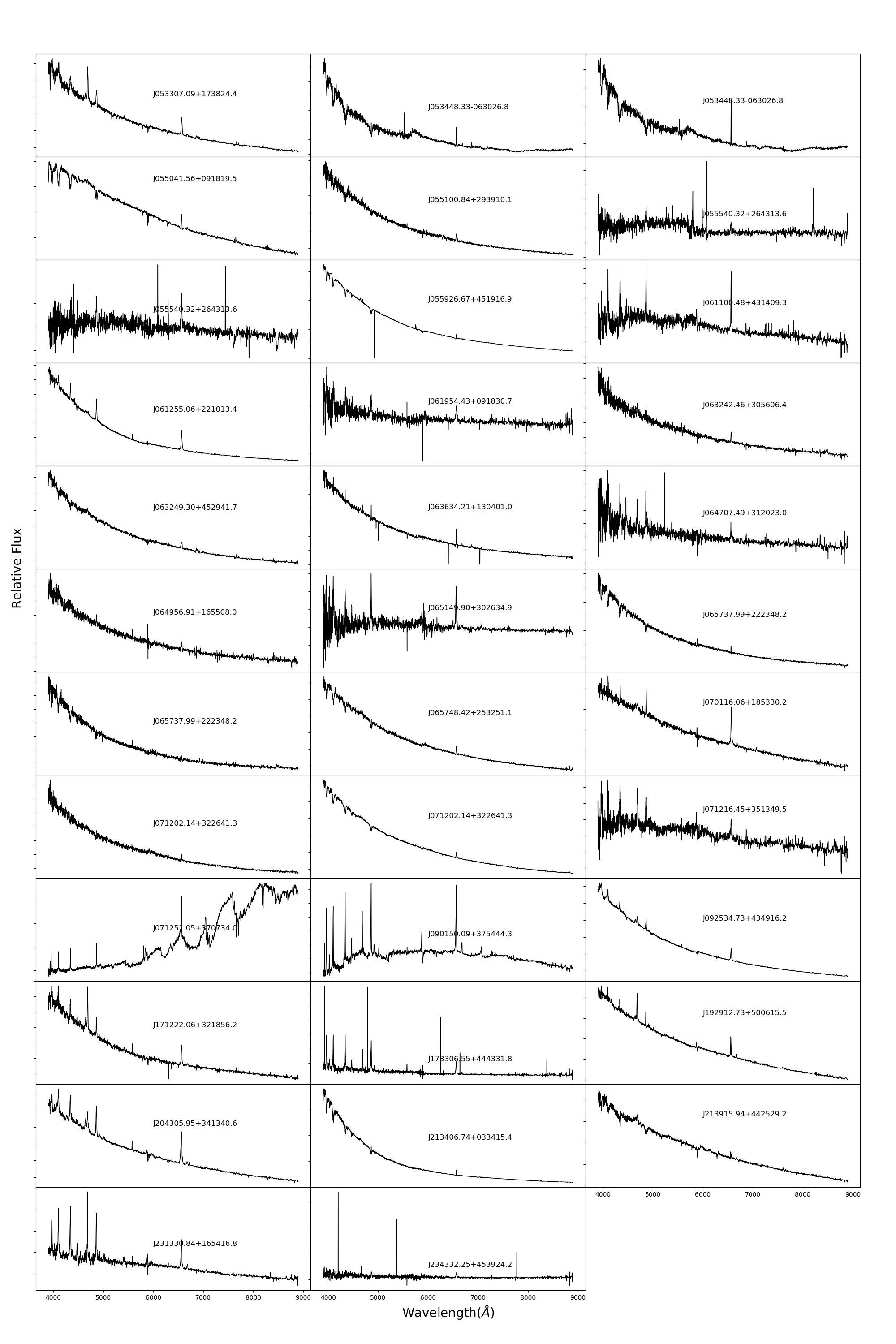}
\caption{Continued.}
\end{figure*}

\begin{figure}[ht!]
\plotone{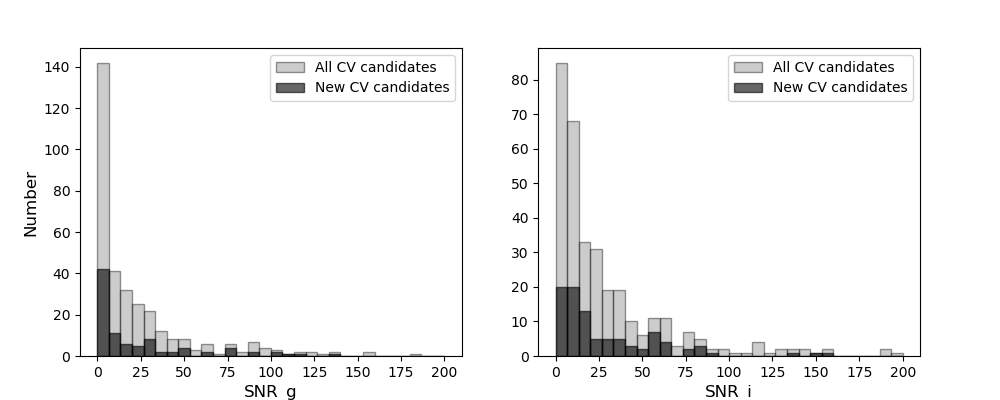}
\caption{The distribution of SNR for CV candidates from LAMOST DR5. The left panel is the SNR distribution of g band, and the right panel is SNR distribution of i band. All the candidates and New ones are repectivelly represented by grey and black. \label{Figure 13}}
\end{figure}

\subsection{Spectra of common CVs in SDSS}
\begin{figure}s
\plotone{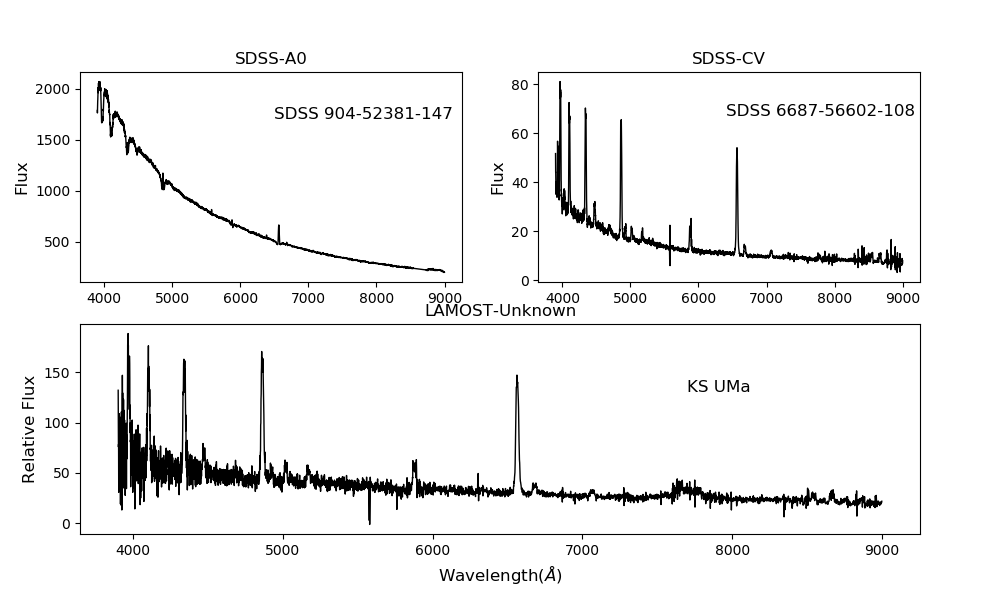}
\plotone{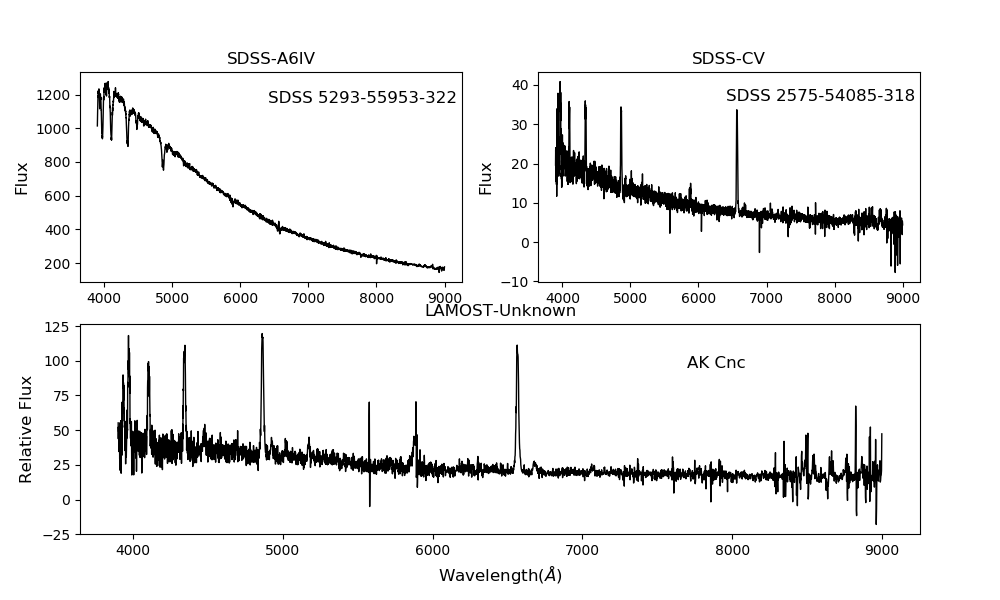}
\caption{The spectra of two CVs from LAMOST and SDSS survey. For each CV, two spectra plotted in the top panels are from SDSS DR14, which are in the states of outburst and quiescence resepctively. The quiescent spectra from LAMOST DR5 are plotted in the bottom panels.\label{Figure 14}}
\end{figure}

We pick up spectra observed in different epochs from SDSS and LAMOST respectively for some CVs. Using a search radius of 5 $\arcsec$, we cross--match the sample of CV candidates from LAMOST DR5 with SDSS DR14 in order to check the spectra of same CVs from SDSS data. Finally, a total of 143 spectra corresponding to 71 CVs are selected from SDSS DR14. Among those spectra, two CVs have spectra both during outburst and in quiescence shown in Figure \ref{Figure 14}. For both CVs, the spectra plotted in the top--left panels which are classified as A--type stars by SDSS DR14 present broad absorption features of Balmer lines, as indicative of its stage of outburst. The spectra with strong emission-lines plotted in the top--right panels indicate that the CVs are at the stage of quiescence, of which the fluxes are almost decreased by one or two orders of magnitude. The amplitudes of the variabilities for the two CVs reveal that both of them are DNs.

\begin{figure}
\plotone{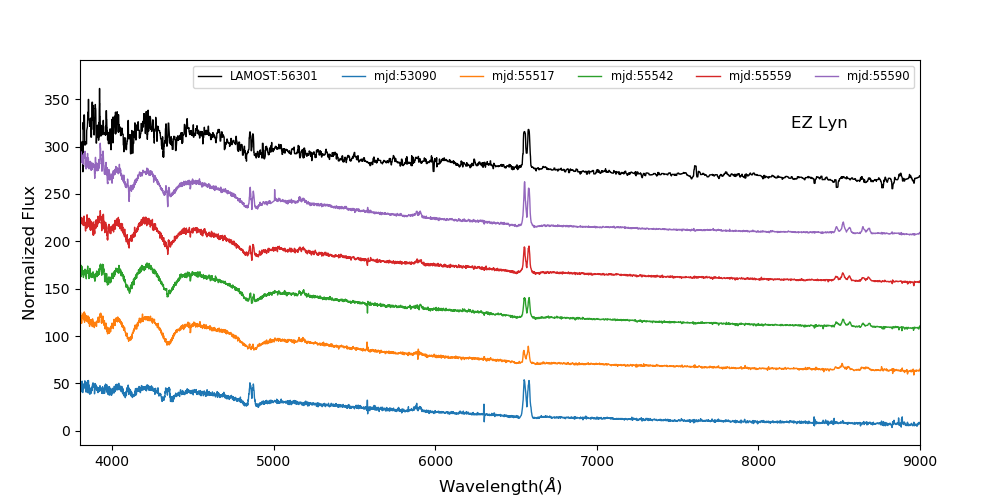}
\caption{Spectra showing the outburst of LAMOST J080434.19+510349.4 (EZ Lyn). This series of spectra are plotted with time. The top spectrum is the latest one which is observed by LAMOST. The bottom five spectra are picked up from SDSS survey. \label{Figure 15}}
\end{figure}

In addition, we also obtain five observations of one CV in SDSS DR14 in Figure \ref{Figure 15}, which is undergoing burst. Four of these spectra are classified as hotter white dwarfs by SDSS pipeline, and another one is classified as CV. The spectrum from LAMOST is also plotted in black in Figure \ref{Figure 15}. From this group of spectra, we can see an obvious change in the emission-line cores of Balmer series, especially in the lower order lines. Moreover, a deep absorption reversals in the emission components indicate that it is a possible high--inclination system.

\subsection{Description of the Catalog}
We provide a catalog of CVs selected from LAMOST DR5 which contains 12 columns. There are 380 spectra of CVs in the catalog, in which 71 spectra are newly discovered. The new discovered CV spectra are marked with an asterisk in the catalog. Table \ref{tab3:catalog} lists 25 records in the catalog as examples, and the complete catalog can be downloaded at http://paperdata.china-vo.org/Whou/CV/CVs\_fitspath\_DR5\_info.csv. The first column of the catalog gives the LAMOST designation for each target, and the second column ID is the identifications of the objects in other literatures if they have been previously identified as CVs. The third and fourth columns are right ascension and declination of the targets respectively. In order to distinguish multiple observations of the same source, we list the the local modified Julian day (lmjd) of the observation date in fifth column. The sixth column gives the subtypes of CVs, which are provided according to the spectral characteristics, including DN, NL, magnetic CV or CV without exact classification whose counts are listed in Table \ref{tab4:subtypes}. The seventh column of Table \ref{tab3:catalog} displays the subtypes of CVs collected from the literatures. Whether the spectra have CIII/NIII and FeII emission lines are indicated in the eighth and ninth columns respectively, while the shapes of H$\alpha$ line profiles are marked as S or D for single--peak or double--peak in the tenth column. In the last two columns, two inverse ratios of line strength H$\beta$/HeII $\lambda$4686 and HeII $\lambda$4686/H$\beta$ are given because they are used as criteria for classifying DN and NL respectively. It is noted that the values of these two ratios might be set to 0, 9999 or -9999 in some cases. The ratios of H$\beta$/HeII $\lambda$4686 and HeII $\lambda$4686/H$\beta$ are both assigned to -9999 in the case where H$\beta$ lines appear in absorption and no HeII $\lambda$4686 emission lines are detected in the spectra. The two ratios are assigned to 0 and 9999 respectively in the case where H$\beta$ lines appear in absorption and HeII $\lambda$4686 emission lines are detected, and on the contrary they are assigned to 9999 and 0 in the case where H$\beta$ lines appear in emission and no HeII $\lambda$4686 emission lines are detected in the spectra. It also needs to be pointed out that 23 spectra in our sample are observed in the test nights, which are not included in the released data. These 23 spectra are specially labelled with '$\sharp$' in the catalog.

\begin{deluxetable*}{cccccccccccc}
\tablecaption{The 25 records of the catalog containing 12 columns \label{tab3:catalog}}
\tablewidth{700pt}
\tabletypesize{\tiny}
\tablehead{
\colhead{Designation} & \colhead{ID\tablenotemark{a}} & \colhead{Ra} & \colhead{Dec} & \colhead{lmjd}
 & \colhead{subtype lamost} & \colhead{subtype literature} & \colhead{CIII/NIII}
 & \colhead{Fe II} & \colhead{line profile (H$\alpha$)} & \colhead{H$\beta$/He4686} & \colhead{He4686/H$\beta$} 
} 
\startdata
J003303.94+380105.4 & CRTS J003304.0+380105 & 8.2664481 & 38.018172 & 57313 & DN & CV & n & y & D & 9999 & 0 \\
J005347.32+405548.5 & SDSS J005347.32+405548.5 & 13.447202 & 40.930155 & 57004 & MCV & CV & y & n & D & 0.55 & 1.82 \\
J013159.86+294922.0 & TT Tri & 22.99943 & 29.82278 & 56993 & DN & CV & n & y & S & 24.76 & 0.04 \\
J013737.21+300248.6 & TX Tri & 24.405066 & 30.046855 & 55918 & DN & DN & n & y & S & 8.18 & 0.12 \\
J020348.61+295925.6 & AI Tri & 30.95255 & 29.990469 & 56976 & MCV & polar & y & n & S & 0.88 & 1.14 \\
J033352.81+332044.5 & PM J03338+3320 & 53.470042 & 33.345695 & 56633 & DN & CV & n & y & D & 9999 & 0 \\
J041329.22+311628.1 & 2MASS J04132921+3116279 & 63.371783 & 31.274486 & 55949 & NL-candidate & CV & y & n & S & 1.87 & 0.54 \\
J051922.90+155434.9 & CRTS J051922.9+155435 & 79.845419 & 15.909696 & 57313 & CV & CV & n & n & S & 27.14 & 0.04 \\
J052658.99+291508.3 & IPHAS J052659.00+291508.4 & 81.745833 & 29.252333 & 55918 & NL-candidate & CV & n & n & S & 1.6 & 0.62 \\
J052832.68+283837.6 & 1RXS J052832.5+283824 & 82.136208 & 28.643778 & 55910 & NL-candidate & polar & y & n & S & 1.78 & 0.56 \\
$^*$J055540.32+264313.6 & - & 88.918024 & 26.720452 & 55899 & CV & none & n & n & S & 9999 & 0 \\
J064734.72+280622.1 & IR Gem & 101.89467 & 28.10615 & 55939 & DN & DN & n & y & S & 5.65 & 0.18 \\
$^*$J064956.91+165508.0 & - & 102.48715 & 16.918906 & 57012 & CV & none & n & n & D & 9999 & 0 \\
$^*$J065149.90+302634.9 & - & 102.95795 & 30.443034 & 55923 & CV & none & n & n & S & 9999 & 0 \\
J084427.11+125232.0 & AC Cnc & 131.1129683 & 12.8755573 & 56645 & CV & CV & n & n & D & 9999 & 0 \\
$^*$J092534.73+434916.2 & - & 141.3947292 & 43.8211917 & 57419 & NL-candidate & star & y & n & S & 1.42 & 0.71 \\
J094431.73+035805.4 & VZ Sex & 146.132225 & 3.9681833 & 57821 & CV & DN & n & n & S & 9999 & 0 \\
J095148.96+340723.5 & RZ LMi & 147.9540125 & 34.1232083 & 56609 & CV & DN & n & n & S & -9999 & -9999 \\
J105430.43+300610.1 & SX LMi & 163.6268151 & 30.1028234 & 55907 & DN & DN & n & y & D & 9999 & 0 \\
J105656.99+494118.2 & CY UMa & 164.23749 & 49.688412 & 56411 & DN & DN & n & y & D & 9.11 & 0.11 \\
$\sharp$J113122.39+432238.5 & MR UMa & 172.8433 & 43.37738 & 57020 & DN & DN & n & y & S & 9999 & 0 \\
J125637.10+263643.2 & GO Com & 194.1546116 & 26.6120105 & 56396 & DN & DN & n & y & S & 9999 & 0 \\
J133941.12+484727.4 & V355 UMa & 204.92135 & 48.790961 & 56394 & DN  & DN & n & y & D & 9999 & 0 \\
J161007.50+035232.7 & V519 Ser & 242.53129 & 3.8757684 & 56359 & CV & polar & n & n & S & 9999 & 0 \\
$^*$J173306.55+444331.8 & - & 263.2773 & 44.7255 & 57870 & NL-candiates & none & y & n & S & 1.12 & 0.89 \\
\enddata
\tablenotetext{a}{The previous identifications of known CVs}
\end{deluxetable*}

\section{Discussion and Summary} \label{sec:summary}
Using the rank-based algorithm in combination with the Random Forest algorithm, we pick out 380 spectra of 245 CVs from LAMOST DR5, among which 71 spectra of 58 CVs are new discoveries. According to the different spectral characteristics of subtypes, a part of the sample are given the subtypes including DN, NL and Magnetic CV. Besides, we also discuss the spectral features of the high--inclination systems and CVs showing the secondaries. We compare the spatial distributions, the magnitude distributions and the Gaia CaMD locus for CVs from LAMOST DR5, SDSS and Catalina surveys. Through cross-matching, we find that some spectra of CVs are categorized as WD or A--type stars in SDSS DR14. More interestingly, two targets that have both spectra in quiescence and during outburst in SDSS DR14, which should be classified as DN according to variation of their luminosities. Besides, we also find that 88 CVs have been observed two or more times in LAMOST survey. We count the frequencies of CVs having at least two spectra and list them in Table \ref{tab5:multi_obs}. The examples of objects which have repeated observations are shown in Figure \ref{Figure 16}. In particular, we note that the object having six observations in Figure \ref{Figure 16} is well known as Feige 24 or FS Cet, which is a close binary system composed of a very hot white dwarf and a M--type main sequence secondary \citep{1976ApJ...210L..87H,2000ApJ...544..423V}. The spectral features of this system resemble those of CVs, while its Balmer emission lines are relatively narrower and generated from the partly ionized atmosphere of M dwarf. From the spectra of some CVs with multiple observations, we can see that the dominated spectral features change from broad absorption lines to strong emission lines, which indicate that they are CVs in the states from the period of quiescence to outburst. In this paper, we mainly focus on the search of CVs in LAMOST dataset, the analysis of spectral characteristics of CVs, and the classification of CV subtypes. Further studies such as analysis of orbital periods will be carried out using the time-domain medium-resolution spectra observed in the phase of LAMOST II, as well as follow--up photometric observations.

\begin{table}[h!]
\renewcommand{\thetable}{\arabic{table}}
\centering
\caption{Numbers of DN, NL, MCV and the remaining CVs without exact subtypes \label{tab4:subtypes}}
\setlength{\tabcolsep}{8pt}
\begin{tabular}{ccccc}
\tablewidth{0pt}
\hline
\hline
\decimals
subtype & DN & NL & MCV & CVs with no subtypes \\
Number of spectra & 134 & 41 & 19 & 186\\
\hline
\end{tabular}
\end{table}

\begin{table}[h!]
\renewcommand{\thetable}{\arabic{table}}
\centering
\caption{Frequency of repeated observations \label{tab5:multi_obs}}
\setlength{\tabcolsep}{8pt}
\begin{tabular}{cccccc}
\tablewidth{0pt}
\hline
\hline
\decimals
Observation Frequency & 2 & 3 & 4 & 5 & 6 \\
Number of Objects & 58 & 19 & 6 & 1 & 4 \\
\hline
\end{tabular}
\end{table}

\begin{figure*}
\plotone{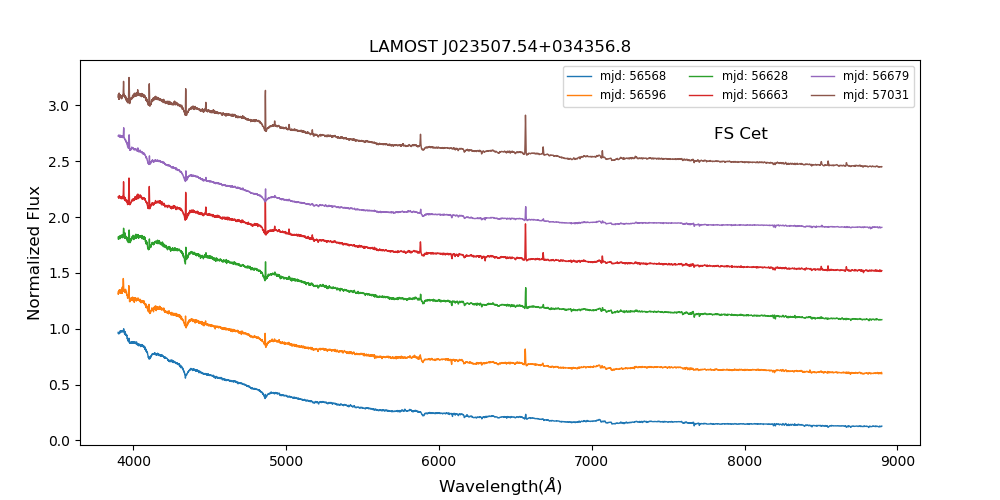}
\plotone{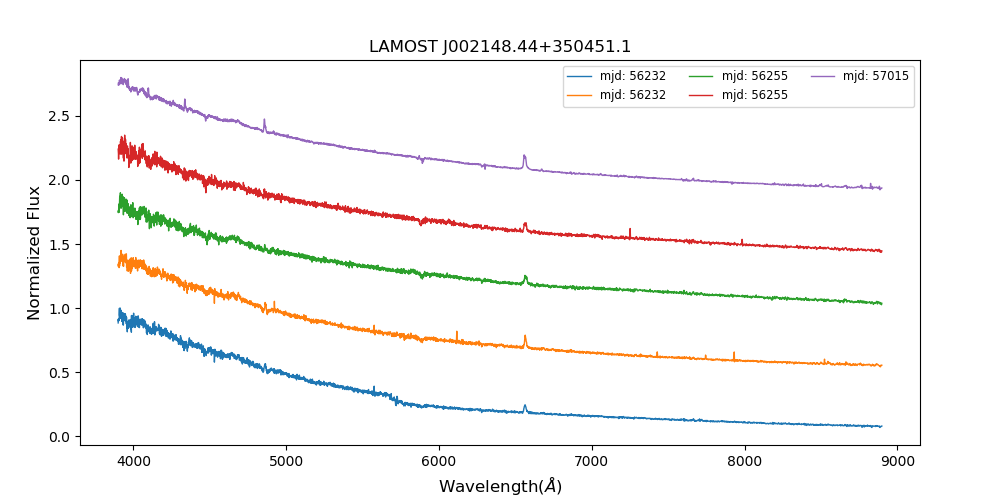}
\plotone{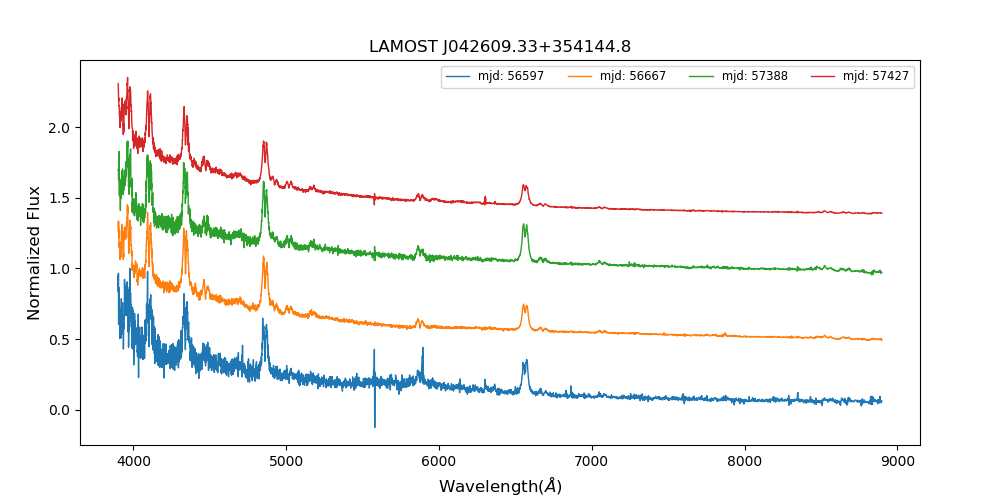}
\caption{Examples of targets which have 2, 3, 4, 5, and 6 observations. In each panel, spectra are plotted by different colors, representing  that they are observed at different epochs.}
\label{Figure 16}
\end{figure*}

\addtocounter{figure}{-1}
\begin{figure*}
\addtocounter{figure}{0}
\plotone{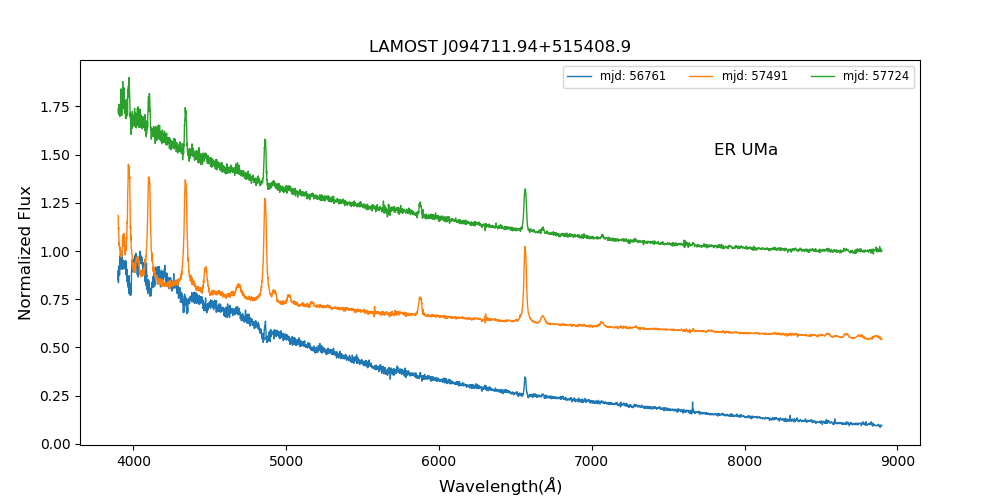}
\plotone{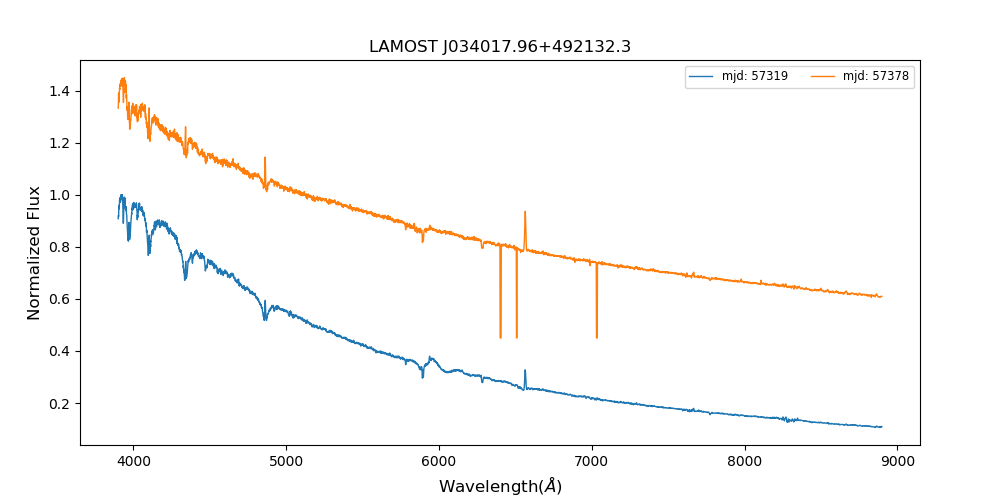}
\caption{Continued.}
\end{figure*}


\acknowledgments

This work is supported by China Scholarship Council, the Joint Research Fund in Astronomy (Grant NO. U1931209) under cooperative agreement between the National Natural Science Foundation of China and Chinese Academy of Sciences. 

Guoshoujing Telescope (the Large Sky Area Multi-Object Fiber Spectroscopic Telescope, LAMOST) is a National Major Scientific Project built by the Chinese Academy of Sciences. Funding for the project has been provided by the National Development and Reform Commission. LAMOST is operated and managed by the National Astronomical Observatories, Chinese Academy of Sciences.

\bibliographystyle{aasjournal}
\bibliography{biblist}

\begin{thebibliography}{}
\expandafter\ifx\csname natexlab\endcsname\relax\def\natexlab#1{#1}\fi
\providecommand{\url}[1]{\href{#1}{#1}}
\providecommand{\dodoi}[1]{doi:~\href{http://doi.org/#1}{\nolinkurl{#1}}}
\providecommand{\doeprint}[1]{\href{http://ascl.net/#1}{\nolinkurl{http://ascl.net/#1}}}
\providecommand{\doarXiv}[1]{\href{https://arxiv.org/abs/#1}{\nolinkurl{https://arxiv.org/abs/#1}}}

\bibitem[{{Breedt} {et~al.}(2014){Breedt}, {G{\"a}nsicke}, {Drake},
  {Rodr{\'{\i}}guez-Gil}, {Parsons}, {Marsh}, {Szkody}, {Schreiber}, \&
  {Djorgovski}}]{2014MNRAS.443.3174B}
{Breedt}, E., {G{\"a}nsicke}, B.~T., {Drake}, A.~J., {et~al.} 2014, \mnras,
  443, 3174, \dodoi{10.1093/mnras/stu1377}

\bibitem[{{Coppejans} {et~al.}(2016){Coppejans}, {K{\"o}rding}, {Knigge},
  {Pretorius}, {Woudt}, {Groot}, {Van Eck}, \& {Drake}}]{2016MNRAS.456.4441C}
{Coppejans}, D.~L., {K{\"o}rding}, E.~G., {Knigge}, C., {et~al.} 2016, \mnras,
  456, 4441, \dodoi{10.1093/mnras/stv2921}

\bibitem[{{Coppejans} {et~al.}(2014){Coppejans}, {Woudt}, {Warner},
  {K{\"o}rding}, {Macfarlane}, {Schurch}, {Kotze}, {Breytenbach}, {Gulbis}, \&
  {Coppejans}}]{2014MNRAS.437..510C}
{Coppejans}, D.~L., {Woudt}, P.~A., {Warner}, B., {et~al.} 2014, \mnras, 437,
  510, \dodoi{10.1093/mnras/stt1900}

\bibitem[{{Cui} {et~al.}(2012){Cui}, {Zhao}, {Chu}, {Li}, {Li}, {Zhang}, {Su},
  {Yao}, {Wang}, {Xing}, {Li}, {Zhu}, {Wang}, {Gu}, {Luo}, {Xu}, {Zhang},
  {Liu}, {Zhang}, {Yang}, {Cao}, {Chen}, {Chen}, {Chen}, {Chen}, {Chu}, {Feng},
  {Gong}, {Hou}, {Hu}, {Hu}, {Hu}, {Jia}, {Jiang}, {Jiang}, {Jiang}, {Jin},
  {Li}, {Li}, {Li}, {Liu}, {Liu}, {Lu}, {Mao}, {Men}, {Qi}, {Qi}, {Shi},
  {Tang}, {Tao}, {Wang}, {Wang}, {Wang}, {Wang}, {Wang}, {Wang}, {Wang},
  {Wang}, {Wang}, {Wang}, {Wang}, {Wang}, {Xu}, {Xu}, {Yang}, {Yu}, {Yuan},
  {Yuan}, {Zhai}, {Zhang}, {Zhang}, {Zhang}, {Zhao}, {Zhou}, {Zhou}, {Zhu}, \&
  {Zou}}]{2012RAA....12.1197C}
{Cui}, X.-Q., {Zhao}, Y.-H., {Chu}, Y.-Q., {et~al.} 2012, Research in Astronomy
  and Astrophysics, 12, 1197, \dodoi{10.1088/1674-4527/12/9/003}

\bibitem[{{Deng} {et~al.}(2012){Deng}, {Newberg}, {Liu}, {Carlin}, {Beers},
  {Chen}, {Chen}, {Christlieb}, {Grillmair}, {Guhathakurta}, {Han}, {Hou},
  {Lee}, {L{\'e}pine}, {Li}, {Liu}, {Pan}, {Sellwood}, {Wang}, {Wang}, {Yang},
  {Yanny}, {Zhang}, {Zhang}, {Zheng}, \& {Zhu}}]{2012RAA....12..735D}
{Deng}, L.-C., {Newberg}, H.~J., {Liu}, C., {et~al.} 2012, Research in
  Astronomy and Astrophysics, 12, 735, \dodoi{10.1088/1674-4527/12/7/003}

\bibitem[{{Downes} {et~al.}(1997){Downes}, {Webbink}, \&
  {Shara}}]{1997PASP..109..345D}
{Downes}, R., {Webbink}, R.~F., \& {Shara}, M.~M. 1997, \pasp, 109, 345,
  \dodoi{10.1086/133900}

\bibitem[{{Downes} \& {Shara}(1993)}]{1993PASP..105..127D}
{Downes}, R.~A., \& {Shara}, M.~M. 1993, \pasp, 105, 127,
  \dodoi{10.1086/133139}

\bibitem[{{Downes} {et~al.}(2001){Downes}, {Webbink}, {Shara}, {Ritter},
  {Kolb}, \& {Duerbeck}}]{2001PASP..113..764D}
{Downes}, R.~A., {Webbink}, R.~F., {Shara}, M.~M., {et~al.} 2001, \pasp, 113,
  764, \dodoi{10.1086/320802}

\bibitem[{{Drake} {et~al.}(2014){Drake}, {G{\"a}nsicke}, {Djorgovski}, {Wils},
  {Mahabal}, {Graham}, {Yang}, {Williams}, {Catelan}, {Prieto}, {Donalek},
  {Larson}, \& {Christensen}}]{2014MNRAS.441.1186D}
{Drake}, A.~J., {G{\"a}nsicke}, B.~T., {Djorgovski}, S.~G., {et~al.} 2014,
  \mnras, 441, 1186, \dodoi{10.1093/mnras/stu639}

\bibitem[{{Du} {et~al.}(2016){Du}, {Luo}, {Yang}, {Hou}, \&
  {Guo}}]{2016PASP..128c4502D}
{Du}, C., {Luo}, A., {Yang}, H., {Hou}, W., \& {Guo}, Y. 2016, \pasp, 128,
  034502, \dodoi{10.1088/1538-3873/128/961/034502}

\bibitem[{{G{\"a}nsicke} {et~al.}(2009){G{\"a}nsicke}, {Dillon}, {Southworth},
  {Thorstensen}, {Rodr{\'{\i}}guez-Gil}, {Aungwerojwit}, {Marsh}, {Szkody},
  {Barros}, {Casares}, {de Martino}, {Groot}, {Hakala}, {Kolb}, {Littlefair},
  {Mart{\'{\i}}nez-Pais}, {Nelemans}, \& {Schreiber}}]{2009MNRAS.397.2170G}
{G{\"a}nsicke}, B.~T., {Dillon}, M., {Southworth}, J., {et~al.} 2009, \mnras,
  397, 2170, \dodoi{10.1111/j.1365-2966.2009.15126.x}

\bibitem[{{Han} {et~al.}(2018){Han}, {Zhang}, {Shi}, {Pi}, {Lu}, {Zhao},
  {Terheide}, \& {Jiang}}]{2018RAA....18...68H}
{Han}, X.~L., {Zhang}, L.-Y., {Shi}, J.-R., {et~al.} 2018, Research in
  Astronomy and Astrophysics, 18, 068, \dodoi{10.1088/1674-4527/18/6/68}

\bibitem[{{Holm}(1976)}]{1976ApJ...210L..87H}
{Holm}, A.~V. 1976, \apjl, 210, L87, \dodoi{10.1086/182310}

\bibitem[{{Hou} {et~al.}(2016){Hou}, {Luo}, {Hu}, {Yang}, {Du}, {Liu}, {Lee},
  {Lin}, {Wang}, {Zhang}, {Cao}, \& {Hou}}]{2016RAA....16..138H}
{Hou}, W., {Luo}, A.-L., {Hu}, J.-Y., {et~al.} 2016, Research in Astronomy and
  Astrophysics, 16, 138, \dodoi{10.1088/1674-4527/16/9/138}

\bibitem[{{Jiang} {et~al.}(2013){Jiang}, {Luo}, {Zhao}, \&
  {Wei}}]{2013MNRAS.430..986J}
{Jiang}, B., {Luo}, A., {Zhao}, Y., \& {Wei}, P. 2013, \mnras, 430, 986,
  \dodoi{10.1093/mnras/sts665}

\bibitem[{{Kogure} \& {Leung}(2007)}]{2007ASSL..342.....K}
{Kogure}, T., \& {Leung}, K.-C., eds. 2007, Astrophysics and Space Science
  Library, Vol. 342, {The Astrophysics of Emission-Line Stars}, 1

\bibitem[{{Li} {et~al.}(2018){Li}, {Luo}, {Du}, {Zuo}, {Wang}, {Zhao}, {Jiang},
  {Zhang}, {Liu}, {Qin}, {Wang}, {Du}, {Guo}, {Wang}, {Han}, {Xiang}, {Huang},
  {Chen}, {Chen}, {Kong}, {Hou}, {Song}, {Wang}, {Wu}, {Zhang}, {Zhang},
  {Wang}, {Cao}, {Hou}, \& {Zhao}}]{2018ApJS..234...31L}
{Li}, Y.-B., {Luo}, A.-L., {Du}, C.-D., {et~al.} 2018, \apjs, 234, 31,
  \dodoi{10.3847/1538-4365/aaa415}

\bibitem[{{Luo} {et~al.}(2012){Luo}, {Zhang}, {Zhao}, {Zhao}, {Cui}, {Li},
  {Chu}, {Shi}, {Wang}, {Zhang}, {Bai}, {Chen}, {Wang}, {Guo}, {Chen}, {Du},
  {Kong}, {Lei}, {Li}, {Song}, {Wu}, {Zhang}, {Zhou}, {Zuo}, {Du}, {He}, {Hou},
  {Dong}, {Li}, {Li}, {Li}, {Song}, {Tian}, {Wang}, {Wu}, {Yang}, {Yuan},
  {Cao}, {Chen}, {Chen}, {Chen}, {Chu}, {Feng}, {Gong}, {Gu}, {Hou}, {Huo},
  {Hu}, {Hu}, {Hu}, {Jia}, {Jiang}, {Jiang}, {Jiang}, {Jin}, {Li}, {Li}, {Li},
  {Li}, {Li}, {Liu}, {Liu}, {Liu}, {Lu}, {Lu}, {Luo}, {Mao}, {Men}, {Ni}, {Qi},
  {Qi}, {Shi}, {Su}, {Sun}, {Su}, {Tang}, {Tao}, {Tu}, {Wang}, {Wang}, {Wang},
  {Wang}, {Wang}, {Wang}, {Wang}, {Wang}, {Wang}, {Wang}, {Wang}, {Wang},
  {Wang}, {Wang}, {Wei}, {Xue}, {Xing}, {Xu}, {Xu}, {Xu}, {Yang}, {Yang},
  {Yao}, {Yu}, {Yuan}, {Zhai}, {Zhang}, {Zhang}, {Zhang}, {Zhang}, {Zhang},
  {Zhang}, {Zhao}, {Zhou}, {Zhu}, {Zhu}, \& {Zou}}]{2012RAA....12.1243L}
{Luo}, A.-L., {Zhang}, H.-T., {Zhao}, Y.-H., {et~al.} 2012, Research in
  Astronomy and Astrophysics, 12, 1243, \dodoi{10.1088/1674-4527/12/9/004}

\bibitem[{{Luo} {et~al.}(2015){Luo}, {Zhao}, {Zhao}, {Deng}, {Liu}, {Jing},
  {Wang}, {Zhang}, {Shi}, {Cui}, {Chu}, {Li}, {Bai}, {Wu}, {Cai}, {Cao}, {Cao},
  {Carlin}, {Chen}, {Chen}, {Chen}, {Chen}, {Chen}, {Chen}, {Chen},
  {Christlieb}, {Chu}, {Cui}, {Dong}, {Du}, {Fan}, {Feng}, {Fu}, {Gao}, {Gong},
  {Gu}, {Guo}, {Han}, {He}, {Hou}, {Hou}, {Hou}, {Hu}, {Hu}, {Hu}, {Huo},
  {Jia}, {Jiang}, {Jiang}, {Jiang}, {Jin}, {Kong}, {Kong}, {Lei}, {Li}, {Li},
  {Li}, {Li}, {Li}, {Li}, {Li}, {Li}, {Li}, {Li}, {Li}, {Li}, {Liang}, {Lin},
  {Liu}, {Liu}, {Liu}, {Liu}, {Lu}, {Luo}, {Mao}, {Newberg}, {Ni}, {Qi}, {Qi},
  {Shen}, {Shi}, {Song}, {Song}, {Su}, {Su}, {Tang}, {Tao}, {Tian}, {Wang},
  {Wang}, {Wang}, {Wang}, {Wang}, {Wang}, {Wang}, {Wang}, {Wang}, {Wang},
  {Wang}, {Wang}, {Wang}, {Wang}, {Wang}, {Wang}, {Wang}, {Wang}, {Wang},
  {Wang}, {Wei}, {Wei}, {Wu}, {Wu}, {Wu}, {Wu}, {Xing}, {Xu}, {Xu}, {Xu},
  {Yan}, {Yang}, {Yang}, {Yang}, {Yang}, {Yao}, {Yu}, {Yuan}, {Yuan}, {Yuan},
  {Yuan}, {Zhai}, {Zhang}, {Zhang}, {Zhang}, {Zhang}, {Zhang}, {Zhang},
  {Zhang}, {Zhang}, {Zhao}, {Zhou}, {Zhou}, {Zhu}, {Zhu}, {Zou}, \&
  {Zuo}}]{2015RAA....15.1095L}
{Luo}, A.-L., {Zhao}, Y.-H., {Zhao}, G., {et~al.} 2015, Research in Astronomy
  and Astrophysics, 15, 1095, \dodoi{10.1088/1674-4527/15/8/002}

\bibitem[{{Luo} {et~al.}(2019)}]{2019luoal}
{Luo}, A.-L., {et~al.} 2019, in prep.

\bibitem[{{Mr{\'o}z} {et~al.}(2015){Mr{\'o}z}, {Udalski}, {Poleski},
  {Pietrukowicz}, {Szyma{\'n}ski}, {Soszy{\'n}ski}, {Wyrzykowski}, {Ulaczyk},
  {Koz{\l}owski}, \& {Skowron}}]{2015AcA....65..313M}
{Mr{\'o}z}, P., {Udalski}, A., {Poleski}, R., {et~al.} 2015, \actaa, 65, 313.
\newblock \doarXiv{1601.02617}

\bibitem[{{Szkody} {et~al.}(2002){Szkody}, {Anderson}, {Ag{\"u}eros},
  {Covarrubias}, {Bentz}, {Hawley}, {Margon}, {Voges}, {Henden}, {Knapp},
  {Vanden Berk}, {Rest}, {Miknaitis}, {Magnier}, {Brinkmann}, {Csabai},
  {Harvanek}, {Hindsley}, {Hennessy}, {Ivezic}, {Kleinman}, {Lamb}, {Long},
  {Newman}, {Neilsen}, {Nichol}, {Nitta}, {Schneider}, {Snedden}, \&
  {York}}]{2002AJ....123..430S}
{Szkody}, P., {Anderson}, S.~F., {Ag{\"u}eros}, M., {et~al.} 2002, \aj, 123,
  430, \dodoi{10.1086/324734}

\bibitem[{{Szkody} {et~al.}(2003){Szkody}, {Fraser}, {Silvestri}, {Henden},
  {Anderson}, {Frith}, {Lawton}, {Owens}, {Raymond}, {Schmidt}, {Wolfe},
  {Bochanski}, {Covey}, {Harris}, {Hawley}, {Knapp}, {Margon}, {Voges},
  {Walkowicz}, {Brinkmann}, \& {Lamb}}]{2003AJ....126.1499S}
{Szkody}, P., {Fraser}, O., {Silvestri}, N., {et~al.} 2003, \aj, 126, 1499,
  \dodoi{10.1086/377346}

\bibitem[{{Szkody} {et~al.}(2004){Szkody}, {Henden}, {Fraser}, {Silvestri},
  {Bochanski}, {Wolfe}, {Ag{\"u}eros}, {Warner}, {Woudt}, {Tramposch}, {Homer},
  {Schmidt}, {Knapp}, {Anderson}, {Covey}, {Harris}, {Hawley}, {Schneider},
  {Voges}, \& {Brinkmann}}]{2004AJ....128.1882S}
{Szkody}, P., {Henden}, A., {Fraser}, O., {et~al.} 2004, \aj, 128, 1882,
  \dodoi{10.1086/423997}

\bibitem[{{Szkody} {et~al.}(2005){Szkody}, {Henden}, {Fraser}, {Silvestri},
  {Schmidt}, {Bochanski}, {Wolfe}, {Ag{\"u}eros}, {Anderson}, {Mannikko},
  {Downes}, {Schneider}, \& {Brinkmann}}]{2005AJ....129.2386S}
{Szkody}, P., {Henden}, A., {Fraser}, O.~J., {et~al.} 2005, \aj, 129, 2386,
  \dodoi{10.1086/429595}

\bibitem[{{Szkody} {et~al.}(2006){Szkody}, {Henden}, {Ag{\"u}eros}, {Anderson},
  {Bochanski}, {Knapp}, {Mannikko}, {Mukadam}, {Silvestri}, {Schmidt},
  {Stephanik}, {Watson}, {West}, {Winget}, {Wolfe}, {Barentine}, {Brinkmann},
  {Brewington}, {Downes}, {Harvanek}, {Kleinman}, {Krzesinski}, {Long},
  {Neilsen}, {Nitta}, {Schneider}, {Snedden}, \& {Voges}}]{2006AJ....131..973S}
{Szkody}, P., {Henden}, A., {Ag{\"u}eros}, M., {et~al.} 2006, \aj, 131, 973,
  \dodoi{10.1086/499308}

\bibitem[{{Szkody} {et~al.}(2007){Szkody}, {Henden}, {Mannikko}, {Mukadam},
  {Schmidt}, {Bochanski}, {Ag{\"u}eros}, {Anderson}, {Silvestri}, {Dahab},
  {Oguri}, {Schneider}, {Shin}, {Strauss}, {Knapp}, \&
  {West}}]{2007AJ....134..185S}
{Szkody}, P., {Henden}, A., {Mannikko}, L., {et~al.} 2007, \aj, 134, 185,
  \dodoi{10.1086/518506}

\bibitem[{{Szkody} {et~al.}(2009){Szkody}, {Anderson}, {Hayden}, {Kronberg},
  {McGurk}, {Riecken}, {Schmidt}, {West}, {G{\"a}nsicke}, {Nebot Gomez-Moran},
  {Schneider}, {Schreiber}, \& {Schwope}}]{2009AJ....137.4011S}
{Szkody}, P., {Anderson}, S.~F., {Hayden}, M., {et~al.} 2009, \aj, 137, 4011,
  \dodoi{10.1088/0004-6256/137/4/4011}

\bibitem[{{Szkody} {et~al.}(2011){Szkody}, {Anderson}, {Brooks},
  {G{\"a}nsicke}, {Kronberg}, {Riecken}, {Ross}, {Schmidt}, {Schneider},
  {Ag{\"u}eros}, {Gomez-Moran}, {Knapp}, {Schreiber}, \&
  {Schwope}}]{2011AJ....142..181S}
{Szkody}, P., {Anderson}, S.~F., {Brooks}, K., {et~al.} 2011, \aj, 142, 181,
  \dodoi{10.1088/0004-6256/142/6/181}

\bibitem[{{Vennes} {et~al.}(2000){Vennes}, {Polomski}, {Lanz}, {Thorstensen},
  {Chayer}, \& {Gull}}]{2000ApJ...544..423V}
{Vennes}, S., {Polomski}, E.~F., {Lanz}, T., {et~al.} 2000, \apj, 544, 423,
  \dodoi{10.1086/317205}

\bibitem[{{Warner}(2003)}]{2003cvs..book.....W}
{Warner}, B. 2003, {Cataclysmic Variable Stars}, 592,
  \dodoi{10.1017/CBO9780511586491}

\bibitem[{{Wenger} {et~al.}(2000){Wenger}, {Ochsenbein}, {Egret}, {Dubois},
  {Bonnarel}, {Borde}, {Genova}, {Jasniewicz}, {Lalo{\"e}}, {Lesteven}, \&
  {Monier}}]{2000A&AS..143....9W}
{Wenger}, M., {Ochsenbein}, F., {Egret}, D., {et~al.} 2000, \aaps, 143, 9,
  \dodoi{10.1051/aas:2000332}

\bibitem[{{Woudt} {et~al.}(2013){Woudt}, {Warner}, {de Bude}, {Macfarlane},
  {Schurch}, \& {Zietsman}}]{2013yCat..74212414W}
{Woudt}, P.~A., {Warner}, B., {de Bude}, D., {et~al.} 2013, VizieR Online Data
  Catalog, 742

\bibitem[{{Zhao} {et~al.}(2012){Zhao}, {Zhao}, {Chu}, {Jing}, \&
  {Deng}}]{2012RAA....12..723Z}
{Zhao}, G., {Zhao}, Y.-H., {Chu}, Y.-Q., {Jing}, Y.-P., \& {Deng}, L.-C. 2012,
  Research in Astronomy and Astrophysics, 12, 723,
  \dodoi{10.1088/1674-4527/12/7/002}

\end{thebibliography}



\end{document}